\documentclass[showpacs,nofootinbib]{revtex4}
\usepackage{mathrsfs,yfonts,amsmath,comment,amssymb,graphicx}
\usepackage[latin1]{inputenc} 
\usepackage[english]{babel}
\DeclareMathAlphabet{\mathpzc}{OT1}{pzc}{m}{it}
\usepackage{hhline}

\newcommand{\up}[1]{$^{(#1)}$}

\newcommand{\ba}{\begin{array}{rcl}}
\newcommand{\ea}{\end{array}}
\newcommand{\be}{\begin{eqnarray}}
\newcommand{\ee}{\end{eqnarray}}
\newcommand{\beq}{\begin{eqnarray}}
\newcommand{\eeq}{\end{eqnarray}}
\newcommand{\bi}{\begin{itemize}}
\newcommand{\ei}{\end{itemize}}

\newcommand{\Z}{\mathbb{Z}}
\newcommand{\pslash}{{p\hspace*{-0.14truecm}\slash}}

\newcommand{\diag}[1]{%
\begin{minipage}[c]{2.5cm}
\includegraphics[width=2.5cm]{#1}
\end{minipage}}

\newcommand{\diagscale}[2]{%
\begin{minipage}[c]{#1}
\includegraphics[width=#1]{#2}
\end{minipage}}

\newcommand{\diagscaleb}[2]{%
\includegraphics[width=#1,origin=B]{#2}
}

\newcommand{\exval}[3]{\left\langle #1 \left| #2 \right| #3 \right\rangle}
\newcommand{\exvalvide}[1]{\exval{0}{#1}{0}}

\newcommand{\lagrange}{{\cal L}}
\newcommand{\Tr}{\mathrm{Tr}}

\newcommand{\partialD}[2]{\frac{\partial #1}{\partial #2}}

\newcommand{\expm}{e^{-\beta\left(E_p-\mu\right)}}

\newcommand{\expmmm}{e^{-3\beta\left(E_p-\mu\right)}}
\newcommand{\expp}{e^{-\beta\left(E_p+\mu\right)}}
\newcommand{\expppp}{e^{-3\beta\left(E_p+\mu\right)}}

\newcommand{\expmT}{e^{-\left(E_p-\mu\right)/T}}

\newcommand{\expmmmT}{e^{-3\left(E_p-\mu\right)/T}}
\newcommand{\exppT}{e^{-\left(E_p+\mu\right)/T}}
\newcommand{\exppppT}{e^{-3\left(E_p+\mu\right)/T}}

\newcommand{\idp}{\int_\Lambda \frac{d^4p}{(2\pi)^4} }
\newcommand{\idpT}{\int_\Lambda \frac{d^3p}{(2\pi)^3} }
\newcommand{\fp}[1]{f(E_{#1} + \mu) }
\newcommand{\fm}[1]{f(E_{#1} - \mu) }
\newcommand{\dtherm}{1-f(E_p - \mu)-f(E_p+\mu) }
\newcommand{\T}{T } 
 
\newcommand{\fsqp}{f^2_P} 
\newcommand{\grdstate}[1]{\left\langle #1 \right\rangle }
\newcommand{\Imag}{\Im m\,}
\newcommand{\Real}{\Re e\,}
\newcommand{\bc}{\begin{center}}
\newcommand{\ec}{\end{center}}
\newcommand{\refp}[1]{(\ref{#1})}

\newcommand{\Tcc}{T^\chi_c}
\newcommand{\Tms}{T^{\sigma-min}_c}

\newcommand{\doublefig}[2]{
\bc
\begin{minipage}{0.75\textwidth}
#1
 \vspace*{2.0truecm} \null \\
#2
\end{minipage}
\ec
}

\begin{document}

\title{Mesonic correlation functions at finite temperature and density in the
Nambu -- Jona - Lasinio model with a Polyakov loop}

\author{H. Hansen\up{a}}\email{hansen@to.infn.it}
\author{W.M.Alberico\up{a}}
\author{A.Beraudo\up{b}}
\author{A.Molinari\up{a}}
\author{M.Nardi\up{a}}
\author{C.Ratti\up{c}}
\affiliation{%
  $(a)$ INFN, Sezione di Torino and Dipartimento di Fisica Teorica,
  University of Torino, via Giuria N.1, 10125 Torino - \textsc{Italy}
}
\affiliation{
  $(b)$ Service de Physique Th\'eorique, CEA Saclay,
  CEA/DSM/SPhT, F-91191, Gif-sur-Yvette - \textsc{France}
}
\affiliation{
  $(c)$ ECT$^*$, 38050 Villazzano (Trento) - \textsc{Italy}
  and INFN, Gruppo Collegato di Trento, via Sommarive, 
  38050 Povo (Trento) - \textsc{Italy}
}

\begin{abstract}
We investigate the properties of scalar and pseudo-scalar mesons at
finite temperature and quark chemical potential in the framework of
the Nambu--Jona-Lasinio (NJL) model coupled to the Polyakov loop (PNJL
model) with the aim of taking into account features of both chiral
symmetry breaking and deconfinement.

The mesonic correlators are obtained by solving the Schwinger--Dyson
equation in the RPA approximation with the Hartree (mean field) quark
propagator at finite temperature and density.

In the phase of broken chiral symmetry a narrower width for the
$\sigma$ meson is obtained with respect to the NJL case; on the other
hand, the pion still behaves as a Goldstone boson.

When chiral symmetry is restored, the pion and $\sigma$ spectral
functions tend to merge. The Mott temperature for the pion is also
computed.
\end{abstract}


\maketitle

\section{Introduction}

Recently, increasing attention has been devoted to study the
modification of particles propagating in a hot or dense medium
\cite{manna,kita}. The possible survival of bound states in the
deconfined phase of QCD
\cite{datta1,shury1,shury2,wong1,wong2,datta2,albe,mocsy} has opened
interesting scenarios for the identification of the relevant degrees
of freedom in the vicinity of the phase transition
\cite{koch,ej,shury3}.  At the same time, renewed interest has arisen
for the study of the $\rho$ meson spectral function in a hot medium
\cite{renk1,renk2,rapp1,rapp2,rho1,rho2}, since precise experimental
data have now become available for this observable \cite{na60}.

In this paper, we focus on the description of light scalar and
pseudo-scalar mesons at finite temperature and quark chemical
potential. Besides lattice calculations \cite{taro,taro2,petr,wis},
high temperature correlators between mesonic current operators can be
studied, starting from the QCD lagrangian, within different
theoretical schemes, like the dimensional reduction \cite{za,lai} or
the Hard Thermal Loop approximation \cite{berry,berry2,mus}.  Actually
both the above approaches rely on a separation of momentum scales
which, strictly speaking, holds only in the weak coupling regime $g\ll
1$. Hence they cannot tell us anything about what happens in the
vicinity of the phase transition.

On the other hand a system close to a phase transition is characterized by
large correlation lengths (infinite in the case of a second order
phase transition). Its behaviour is mainly driven by the symmetries of
the lagrangian, rather than by the details of the microscopic
interactions.  In this critical regime of temperatures and densities
our investigation of meson properties is then performed in the
framework of an effective model of QCD, namely a modified Nambu
Jona-Lasinio model including Polyakov loop dynamics (referred to as
PNJL model) ~\cite{Meisinger:1995ih,Meisinger:2001cq,Fukushima:2003fw,
Mocsy:2003qw,Megias:2004hj,Ratti:2005jh,Ratti:2006gh,mus2} .

Models of the Nambu and Jona-Lasinio (NJL) type \cite{NJL61} have a
long history and have been extensively used to describe the dynamics
and thermodynamics of the lightest hadrons
\cite{Hatsuda:1985eb,Bernard:1987im,Bernard:1987ir,Jaminon:1989wp,VW91,Klevansky:1992qe,Lutz:1992dv,HK94,Bernard:1990ye,Ri97}. Such schematic models offer a
simple and practical illustration of the basic mechanisms that drive
the spontaneous breaking of chiral symmetry, a key feature of QCD in
its low-temperature, low-density phase.

In first approximation the behavior of a system ruled by QCD is
governed by the symmetry properties of the Lagrangian, namely the
(approximate) global symmetry $SU_L(N_f) \times SU_R(N_f)$, which is
spontaneously broken to $SU_V(N_f)$ and the (exact) $SU_c(N_c)$ local
color symmetry.
Indeed in the NJL model the mass of a constituent quark is directly
related to the chiral condensate, which is the order parameter of the
chiral phase transition and, hence, is non-vanishing at zero
temperature and density. Here the system lives in the phase of
spontaneously broken chiral symmetry: the strong interaction, by
polarizing the vacuum and turning it into a condensate of
quark-antiquark pairs, transforms an initially point-like quark with
its small bare mass $m_0$ into a massive quasiparticle with a finite
size.
Despite their widespread use, NJL models suffer a major shortcoming:
the reduction to global (rather than local) colour symmetry prevents
quark confinement.

On the other hand, in a non-abelian pure gauge theory, the Polyakov
loop serves as an order parameter for the transition from the low
temperature, $\mathbb{Z}_{N_c}$ symmetric, confined phase (the active
degrees of freedom being color-singlet states, the glueballs), to the
high temperature, deconfined phase (the active degrees of freedom
being colored gluons), characterized by the spontaneous breaking of
the $\mathbb{Z}_{N_c}$ (center of $SU_c(N_c)$) symmetry.

With the introduction of dynamical quarks, this symmetry breaking
pattern is no longer exact: nevertheless it is still possible to
distinguish a hadronic (confined) phase from a QGP (deconfined) one.

In the PNJL model quarks are coupled simultaneously to the chiral
condensate and to the Polyakov loop: the model includes features of
both chiral and $\Z_{N_c}$ symmetry breaking. The model has proven to
be successful in reproducing lattice data concerning QCD
thermodynamics~ \cite{Ratti:2005jh}. The coupling to the Polyakov
loop, resulting in a suppression of the unwanted quark contributions
to the thermodynamics below the critical temperature, plays a
fundamental role for this purpose.

It is therefore natural to investigate the predictions of the PNJL
model for what concerns mesonic properties. Since the ``classic'' NJL
model lacks confinement, the $\sigma$ meson for example can
unphysically decay into a $\bar{q} q$ pair even in the vacuum:
indeed this process is energetically allowed and there is no mechanism
which can prevent it. As a consequence, the $\sigma$ meson shows, in
the NJL model, an unphysical width corresponding to this
process. One of our goals is to check whether the coupling of
quarks to the Polyakov loop is able to cure this problem, thus
preventing the decay of the $\sigma$ meson into a $\bar{q}q$
pair. Accordingly, particular emphasis will be given in our work to the
$\sigma$ spectral function.

We compute the mesonic correlation functions in ring approximation
(i.e. RPA, if one neglects the antisymmetrisation) with quark
propagator evaluated at the Hartree mean field level.  The properties
of mesons at finite temperature and chemical potential are finally
extracted from these correlation functions.  We restrict ourselves to
the scalar-pseudoscalar sectors and discuss the impact of the Polyakov
loop on the mesonic properties and the differences between NJL and
PNJL models.  Due to the simplicity of the model where dynamical
gluonic degrees of freedom are absent, no true mechanism of
confinement is found (we will show that for the $\sigma$ meson the
decay channel $\sigma\rightarrow q\bar{q}$ is still open also below
$T_c$).

Our paper is organized as follows: in Section~\ref{model} we briefly
review the main features of the PNJL model, how quarks are coupled to
the Polyakov loop, our parameter choice and some results obtained in
Ref.~\cite{Ratti:2005jh} which are relevant to our work. In
Sections~\ref{correlators} and~\ref{spectral} we address the study of
correlators of current operators carrying the quantum numbers of
physical mesons, and the corresponding mesonic spectral functions and
propagators; we obtain the relevant formulas both in the NJL and in
the PNJL cases, and discuss the main differences between the two
models.  Our numerical results concerning the mesonic masses and
spectral functions are discussed in
Section~\ref{numerical}. Particular attention is again focused on the
NJL/PNJL comparison. Final discussions and conclusions are presented
in Section~\ref{discussion}.

\section{The model \label{model}}
\subsection{Nambu -- Jona - Lasinio model}
Motivated by the symmetries of QCD, we use the NJL model (see
\cite{VW91,Klevansky:1992qe,HK94,Buballa:2003qv} for review papers) for the
description of the coupling between quarks and the chiral
condensate in the scalar-pseudoscalar sector. We will use a two flavor
model, with a degenerate mass matrix for quarks.
The associated Lagrangian reads:
\beq
 \lagrange_{NJL} &=& \bar q ( i \gamma^\mu \partial_\mu - \hat m) q 
   + {G_1}  \left[ {\left( \bar q q \right)}^2 
   + {\left( \bar q i \gamma_5 \vec\tau q \right)}^2 \right] 
\label{NJL}
\eeq
In the above $\bar q = (\bar u, \bar d)$, $\hat m = diag(m_u,m_d)$,
with $m_u=m_d \equiv m_0$ (we keep the isospin symmetry); finally
$\tau^a~(a=1,2,3)$ are $SU_f(2)$ Pauli matrices acting in flavor
space.  As it is well known, this Lagrangian is invariant under a
global -- and not local -- color symmetry $SU(N_c=3)$ and lacks the
confinement feature of QCD.  It also satisfies the chiral $SU_L(2)
\times SU_R(2)$ symmetry if $\hat m = 0$ while $\hat m \neq 0 $
implies an explicit (but small) chiral symmetry breaking from $SU_L(2)
\times SU_R(2)$ to $SU_f(2)$ which is still exact, due to the choice
$m_u=m_d\equiv m_0$.

The parameters entering into Eq.~(\ref{NJL}) are usually fixed to
reproduce the mass and decay constant of the pion as well as the
chiral condensate.  The parameters we use are given in Table
\ref{NJLparam}, together with the calculated physical quantities
chosen to fix the parameters. The Hartree quark mass (or constituent
quark mass) is $m = 325$ MeV and the pion decay constant and mass are
obtained within a Hartree + RPA calculation.

\newcommand{\quarkdensity}{$|\langle{\bar \psi}_u\psi_u\rangle|^{1/3}$}
\begin{table}
\begin{center}
\begin{tabular}{||c|c|c||c|c|c||}
\hhline{|t:===:t:===:t|}
$\Lambda$ [GeV]      &  $G_1$ [GeV$^{-2}$]  &  $m_0$ [MeV]    & 
\phantom{\bigg(}\quarkdensity\ [MeV] & $f_{\pi}$ [MeV]      & $m_{\pi}$ [MeV] 
\\
\hline
\phantom{\bigg(} 0.651 &               5.04 &           5.5 &
                   251 &               92.3 &         139.3 \\
\hhline{|b:===:b:===:b|}
\end{tabular}
\caption{Parameter set for the NJL Lagrangian given in Eq.~(\ref{NJL})
  and the physical quantities chosen to fix the parameters.}
\label{NJLparam}
\end{center}
\end{table}

\subsection{Pure gauge sector}
In this Section, following the arguments given in \cite{Pisa1,Pisa2},
we discuss how the deconfinement phase transition in a pure $SU(N_c)$
gauge theory can be conveniently described through the introduction of
an effective potential for the complex Polyakov loop field, which we
define in the following.

Since we want to study the $SU(N_c)$ phase structure, first of all
an appropriate order parameter has to be defined.
For this purpose the Polyakov line
\be
L
\left(\vec{x}\right)&\equiv&\mathcal{P}\exp\left[i\int_{0}^{\beta}d\tau\,
A_4\left(\vec{x},\tau\right)\right]\label{eq:defpoly}
\ee
is introduced. In the above, $A_4 = i A^0$ is the temporal component of
the Euclidean gauge field $(\vec{A}, A_4)$, in which the strong
coupling constant $g_S$ has been absorbed, $\mathcal{P}$ denotes path
ordering and the usual notation $\beta = 1/T$ has been introduced with
the Boltzmann constant set to one ($k_B = 1$).

When the theory is regularized on the lattice, 
the Polyakov loop,
\beq
l(\vec{x})=\frac{1}{N_c} \Tr {L(\vec{x})} ,
\eeq
is a color singlet under $SU(N_c)$, but transforms non-trivially, like
a field of charge one, under $\Z_{N_c}$. Its thermal expectation value
is then chosen as an order parameter for the deconfinement phase
transition \cite{Poly1,'thooft,Svet}.
In fact, in the usual physical interpretation \cite{McLerr,Rothe},
$\langle l(\vec{x})\rangle$ is related to the change of free energy
occurring when a heavy color source in the fundamental representation
is added to the system. One has:
\beq
\langle l(\vec{x})\rangle=e^{-\beta\Delta F_Q(\vec{x})}.
\label{criterion}
\eeq
In the $\Z_{N_c}$ symmetric phase, $\langle l(\vec{x})\rangle=0$,
implying that an infinite amount of free energy is required to add an isolated 
heavy quark to the system: in this phase color is confined.

Phase transitions are usually characterized by large correlation
lengths, i.e. much larger than the average distance between the
elementary degrees of freedom of the system. Effective field theories
then turn out to be a useful tool to describe a system near a phase
transition.  In particular, in the usual Landau-Ginzburg approach, the
order parameter is viewed as a field variable and for the latter an
effective potential is built, respecting the symmetries of the original
lagrangian. 
In the case of the $SU(3)$ gauge theory, the Polyakov line $L(\vec{x})$
gets replaced by its gauge covariant average over a finite region of
space, denoted as $\langle\!\langle L(\vec{x})\rangle\!\rangle$
\cite{Pisa1}. Note that $\langle\!\langle L(\vec{x})\rangle\!\rangle$ in
general is not a $SU(N_c)$ matrix.
The Polyakov loop field:
\beq
\Phi(\vec x)\equiv\langle\!\langle l(\vec{x})\rangle\!\rangle=\frac 1 {N_c}\,\Tr_c \, \langle\!\langle L(\vec x)\rangle\!\rangle
\eeq
is then introduced.

Following \cite{Pisa1,Pisa2,Ratti:2005jh}, we define an effective
potential for the (complex) $\Phi$ field, which is conveniently chosen
to reproduce, at the mean field level, results obtained in lattice
calculations. In this approximation one simply sets the Polyakov loop
field $\Phi(\vec{x})$ equal to its expectation value $\Phi=$const.,
which minimizes the potential
\beq
\frac{\mathcal{U}\left(\Phi,\bar\Phi;T\right)}{T^4} =-\frac{b_2\left(T\right)}{2}\bar\Phi \Phi-
\frac{b_3}{6}\left(\Phi^3+
{\bar\Phi}^3\right)+ \frac{b_4}{4}\left(\bar\Phi \Phi\right)^2\;,
\label{Ueff}
\eeq
where
\beq
b_2\left(T\right)=a_0+a_1\left(\frac{T_0}{T}\right)+a_2\left(\frac{T_0}{T}
\right)^2+a_3\left(\frac{T_0}{T}\right)^3.
\eeq
A precision fit of the coefficients $a_i,~b_i$ has been performed in
Ref. \cite{Ratti:2005jh} to reproduce some pure-gauge lattice data.
The results are reported in Table \ref{paramPG}.  These parameters
have been fixed to reproduce the lattice data for both the expectation
value of the Polyakov loop \cite{latticePL} and some thermodynamic
quantities~\cite{boyd1996}. 
The parameter $T_0$ is the critical
temperature for the deconfinement phase transition, fixed to $270$ MeV
according to pure gauge lattice findings.
With the present choice of the parameters, $\Phi$ and $\bar\Phi$
are never larger than one in the pure gauge sector. The lattice data
in Ref.  \cite{latticePL} show that for large temperatures the
Polyakov loop exceed one, a value which is reached asymptotically from
above. This feature cannot be reproduced in the absence of radiative
corrections: therefore, at the mean field level, it is consistent to
have $\Phi$ and $\bar\Phi$ always smaller than one. In any case, the
range of applicability of our model is limited to temperatures $T\leq
2.5~T_c$ (see the discussion at the end of the next section) and for
these temperatures there is good agreement between our results and the
lattice data for $\Phi$.  
\begin{table}
\begin{center}
\begin{tabular}{|c|c|c|c|c|c|}
\hhline{|t:===:t:===:t|}
$a_0$ & $a_1$ & $a_2$ & $a_3$ & $b_3$ & $b_4$ \\
\hline
6.75  & -1.95 & 2.625 & -7.44 & 0.75  &   7.5 \\
\hhline{|t:===:t:===:t|}
\end{tabular}
\caption{Parameters for the effective potential in the pure gauge sector 
(Eq.~(\ref{Ueff})).}
\label{paramPG}
\end{center}
\end{table}

\begin{figure}
\doublefig{ $T=0.26$ GeV$<T_0$ \\
``Color confinement'' \\
$\grdstate{\Phi} = 0 \longrightarrow$ No breaking of $\mathbb{Z}_3$  \\
\includegraphics[width=0.95\textwidth]{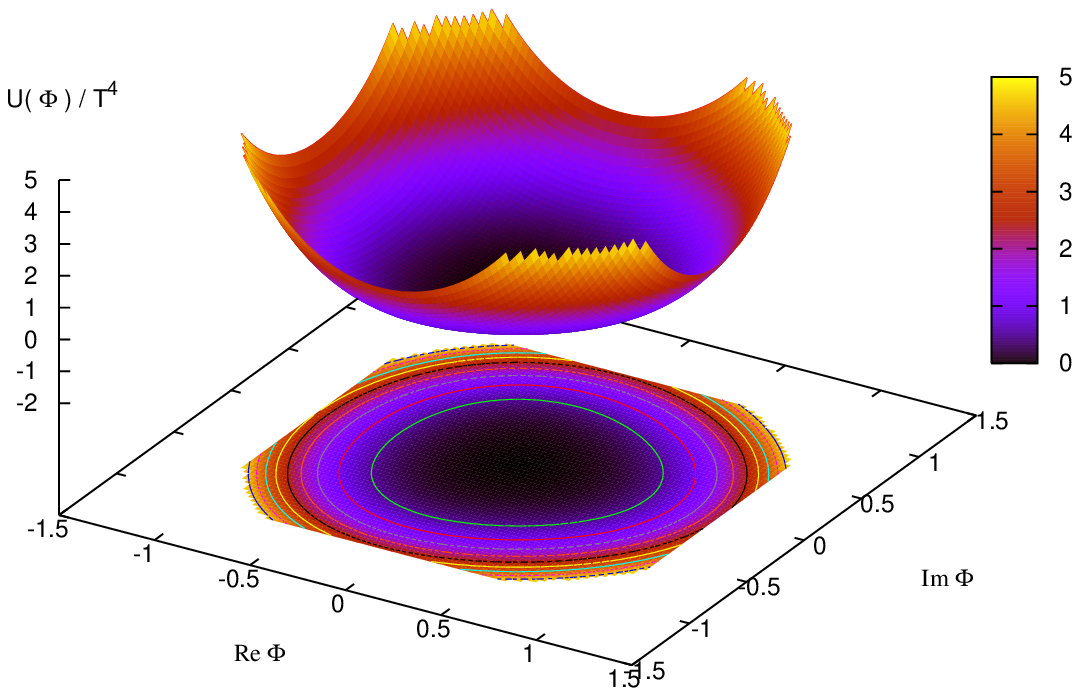} } { $T=1$ GeV $>T_0$ \\
``Color deconfinement'' \\
$\grdstate{\Phi} \neq 0 \longrightarrow$ breaking of $\mathbb{Z}_3$ \\
\includegraphics[width=0.95\textwidth]{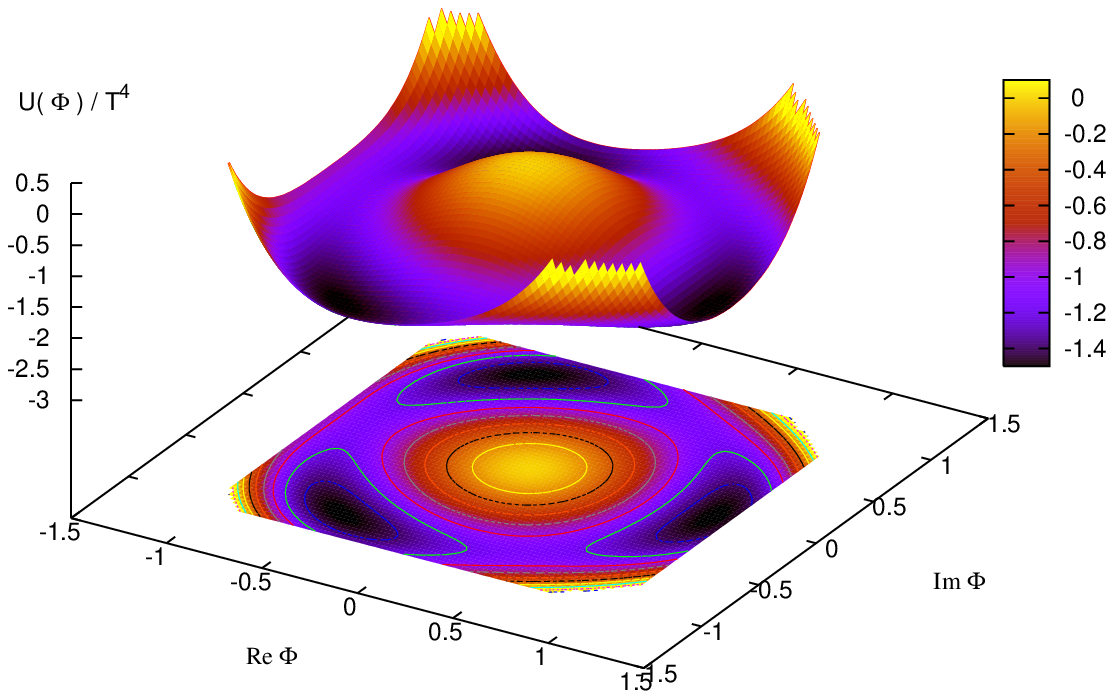} }
\caption{Effective potential in the pure gauge sector (Eq.~(\ref{Ueff}))
for two characteristic temperatures, below and above
the critical temperature $T_0$. One can see three minima appearing
above $T_0$.}
\label{fig:PGEffPot}
\end{figure}

The effective potential presents the feature of a phase transition
from color confinement ($T<T_0$, { the minimum of the effective
potential being at $\Phi=0$}) to color deconfinement ($T>T_0$, { the
minima of the effective potential occurring at $\Phi \neq 0$}) as it
can be seen from { Fig.} \ref{fig:PGEffPot}.  The potential possesses
the $\Z_3$ symmetry and one can see that, above $T_0$, it presents
three minima ($\Z_3$ symmetric), showing a spontaneous symmetry
breaking.

\subsection{Coupling between quarks and the gauge sector: the PNJL model}

In the presence of dynamical quarks the $\Z_3$ symmetry is explicitly
broken. One cannot rigorously talk of a phase transition, but the
expectation value of the Polyakov loop still serves as an indicator
for the crossover between the phase where color confinement occurs
($\Phi \longrightarrow 0$) and the one where color is deconfined
($\Phi \longrightarrow 1$).

The PNJL model attempts to describe in a simple way the two
characteristic phenomena of QCD, namely deconfinement and chiral
symmetry breaking.

In order to describe the coupling of quarks to the chiral condensate, we
start from an NJL description of quarks (global $SU_c(3)$ symmetric
point-like interaction), coupled in a minimal way to the Polyakov loop, via 
the following
Lagrangian (\cite{Ratti:2005jh})\footnote{We use here the original
Lagrangian of Ref.~\cite{Ratti:2005jh}, with a complex Polyakov loop
effective field, which implies that at $\mu\ne 0$ the expectation
values of $\Phi$ and $\bar\Phi $ are different. A
different choice can be motivated~\cite{Weise-privatecommunication} but we have checked that the
calculations of the present work are not sensitive to this
feature.  }:
\be
\mathcal{L}_{PNJL}=\bar{q}\left(i\gamma_{\mu}D^{\mu}-\hat{m}_0
\right)q+
G_1 \left[\left(\bar{q}q\right)^2+\left(\bar{q}i\gamma_5
\vec{\tau}q
\right)^2\right]
-\mathcal{U}\left(\Phi[A],\bar\Phi[A];T\right),
\label{lagr:PNJL}
\ee
where the covariant derivative reads $D^{\mu}=\partial^\mu-i A^\mu$
and $A^\mu=\delta^{\mu}_{0}A^0$ (Polyakov gauge), with $A^0 = -iA_4$.  The strong
coupling constant $g_S$ is absorbed in the definition of $A^\mu(x) =
g_S {\cal A}^\mu_a(x)\frac{\lambda_a}{2}$ where ${\cal A}^\mu_a$ is
the gauge field ($SU_c(3)$) and $\lambda_a$ are the Gell--Mann matrices.
We notice explicitly that at $T=0$ the Polyakov loop and the quark sector decouple.

In order to address the finite density case, it turns out to be useful
to introduce the following effective Lagrangian:
\beq
\mathcal{L}'_{PNJL}=\mathcal{L}_{PNJL}+\mu\bar{q}\gamma^0q\;,\label{eq:leff}
\eeq
which leads to the customary grand canonical Hamiltonian.
In the above the chemical potential term accounts for baryon
number conservation which, in the grand canonical ensemble, is not
imposed exactly, but only through its expectation value.
Let us comment here the range of applicability of the PNJL model.
As already stated in Ref.~\cite{Ratti:2005jh}, in the PNJL model the
gluon dynamics is reduced to a chiral-point coupling between quarks
together with a simple static background field representing the
Polyakov loop. This picture cannot be expected to work outside a
limited range of temperatures.  At large temperatures transverse
gluons are known to be thermodynamically active degrees of freedom:
they are not taken into account in the PNJL model.
Hence based on the conclusions drawn in \cite{Meisinger:2003id}
according to which transverse gluons start to contribute significantly
for $T>2.5\,T_c$, we can assume that the range of applicability of the
model is limited roughly to $T\leq (2-3)T_c$. 

\subsection{Field equations}

\subsubsection{Hartree approximation}
In this Section we derive the \emph{gap equation} in the Hartree
approximation, whose solution provides the self-consistent PNJL mass
of the dressed quark.

We start from the effective lagrangian given in 
Eq.~(\ref{eq:leff}). The imaginary time formalism is employed. 
One defines the vertices $\Gamma_M$, where $M=\{S,P\}$, in the scalar
($\Gamma_S \equiv \mathbb{I}$) and pseudo-scalar ($\Gamma_P^a \equiv i \gamma_5 \tau^a$) channel.
The diagrammatic Hartree equation reads: \\
\be
\diagscale{1.2cm}{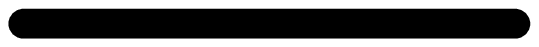} = \diagscale{1.2cm}{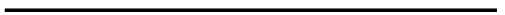} + \diagscale{1.2cm}{g0} \diagscale{1.2cm}{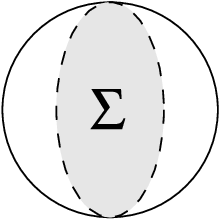} \diagscale{1.2cm}{g}
 = \diagscale{1.2cm}{g0} + \diagscaleb{3.6cm}{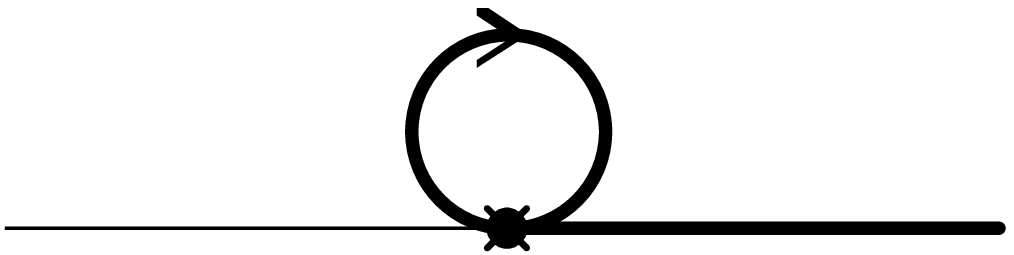}
\ee
where the thin line denotes the free propagator in the constant (we work in the mean field) background 
field $A_4$: $S_0(p) =
\diagscale{0.7cm}{g0} = -(\pslash - m_0+\gamma^0(\mu-iA_4))^{-1}$, the thick line the
Hartree propagator $S(p) = \diagscale{0.7cm}{g} = -(\pslash - m+\gamma^0(\mu-iA_4))^{-1}$,
the cross ($\diagscale{0.25truecm}{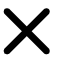}$) the vertex
$\Gamma_M$ and the dot ($\diagscale{0.25truecm}{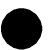}$) represents $2G_1$, the
coupling constant in the scalar-pseudoscalar channel (indeed due to parity
invariance only the scalar vertex contributes). \\
Besides, $\diagscale{1.5cm}{Sigma} = \diagscale{1cm}{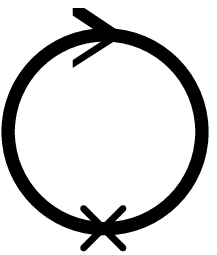} $ is the
Hartree self-energy and $m \equiv m_0 + \Sigma$. The Hartree equation
then reads:
\beq
  m-m_0=2G_1 T\ \Tr\sum_{n=-\infty}^{+\infty}\int_\Lambda\frac{\mathrm{d}^3p}{\left(2\pi\right)^3}\frac{-1}{\pslash - m+\gamma^0(\mu-iA_4)}
\label{eq:Hartree_brut}
\eeq
In all the above formulas, $p_0 = i \omega_n$ and $\omega_n =(2n+1)\pi
T$ is the Matsubara frequency for a fermion; the trace is taken over
color, Dirac and flavor indices. The symbol $\int_\Lambda$ denotes the
three dimensional momentum regularisation; we use an ultraviolet
cut-off $\Lambda$ for both the zero and the finite temperature
contributions. Our choice is motivated by our wish to discuss mesonic
properties driven by chiral symmetry considerations, a feature not
well described if one only regularizes the $T=0$ part (in particular
in the vector sector the Weinberg sum rule is not well satisfied).
Through a convenient gauge transformation of the Polyakov line,
 the background field $A_4$ in Eq.~(\ref{eq:Hartree_brut}) can always be put
in a diagonal form. This allows one to straightforwardly perform the sum over
the Matsubara frequencies yielding (see also section \ref{section:finiteT}):
\beq
  m-m_0=2G_1 N_f\sum_{i=1}^{N_c}\int_\Lambda\frac{\mathrm{d}^3p}{\left(2\pi\right)^3}\frac{2m}{E_p}[1-f(E_p-\mu+iA_4^{ii})-f(E_p+\mu-iA_4^{ii})]\,.
\label{eq:Andrea1}
\eeq
By introducing the modified distribution functions\footnote{We will
explicitly derive these quantities and their role in
Sec.~\ref{section:finiteT}.}  $f^+_\Phi$ and $f^-_\Phi$, here derived
for $N_c = 3$ (with the usual notation $\beta = 1/T$):
\be
  f^+_\Phi(E_p) 
         & =&\frac{ \left( \Phi + 2\bar\Phi \expm \right) \expm + \expmmm }
  {1 + 3\left( \Phi + \bar\Phi \expm \right) \expm + \expmmm} \label{fpPhi}\\
  f^-_\Phi(E_p) 
         & =&\frac{ \left( \bar\Phi + 2\Phi \expp \right) \expp + \expppp }
  {1 + 3\left( \Phi + \bar\Phi \expp \right) \expp + \expppp}, \label{fmPhi}
\ee
the gap equation reads:
\beq
  m-m_0=2G_1 N_f N_c\int_\Lambda\frac{\mathrm{d}^3p}{\left(2\pi\right)^3}\frac{2m}{E_p}[1-f^+_\Phi(E_p)-f^-_\Phi(E_p)]\,.
\label{eq:Andrea4}
\eeq
The latter is valid for any $N_c$ providing one uses the corresponding  $f^{+,-}_\Phi$. 
Notice that Eq.~\refp{eq:Hartree_brut}, after computing the trace on Dirac and
isospin indices, can be viewed as a generalization of the
corresponding zero temperature and density NJL gap equation
\be
  m-m_0 &=& 8iG_1m N_c N_f \idp \frac{1}{p^2-m^2} ,
\label{eq:Hartree}
\ee
after adopting the following symbolic replacements:
\be
  p = (p_0, \vec p) & \rightarrow & (i \omega_n + \mu - i A_4, \vec p) 
\label{replacement1} \\
  i\idp        & \rightarrow & - T \frac{1}{N_c} \Tr_c \sum_n \idpT ,
\label{replacement2}
\ee
\subsubsection{Grand potential at finite temperature and density in
  Hartree approximation}

The usual techniques \cite{Klevansky:1992qe,Schwarz:1999dj} can be
 used to obtain the PNJL grand potential from the Hartree
propagator (see \cite{Ratti:2005jh}):
\be
\Omega = \Omega(\Phi, \bar\Phi, m ; T, \mu) &=&{\cal U}\left(
\Phi,\bar{\Phi},T\right)+\frac{(m - m_0)^2}{4G_1}
- 2 N_c N_f\int_\Lambda\frac{\mathrm{d}^3p}{\left(2\pi\right)^3}\,{E_p} \nonumber \\
&& \hspace{-2cm} -2N_f\,T\int_\Lambda\frac{\mathrm{d}^3p}{\left(2\pi\right)^3}
\left\{\mathrm{Tr}_c\ln\left[1+ L^\dagger \mathrm{e}^{-\left(E_p-\mu
\right)/T}\right]
+\mathrm{Tr}_c\ln\left[1+ L
 \mathrm{e}^{-\left(E_p+\mu\right)/T}\right]
\right\}~. 
\nonumber\\
\label{omega} 
\ee
In the above formula
$E_p=\sqrt{\vec{p}\,^2+m^2}$ is the Hartree single quasi-particle energy (which
includes the constituent quark mass). We then define $z^{+,-}_\Phi$ and compute 
them for $N_c = 3$: 
\be
z^+_\Phi \equiv \mathrm{Tr}_c\ln\left[1+ L^\dagger \mathrm{e}^{-\left(E_p-\mu\right)/T}\right] &=&
     \ln\left\{       1 + 3\left( \bar\Phi + \Phi \expmT \right) \expmT
 + \expmmmT \right\} \ \label{eq:termo1}\label{zplus}
\\
z^-_\Phi \equiv \mathrm{Tr}_c\ln\left[1+ L \mathrm{e}^{-\left(E_p+\mu\right)/T}\right] &=& 
       \ln\left\{     1 + 3\left( \Phi + \bar\Phi \exppT \right) \exppT
 + \exppppT \right\}~.\ \label{eq:termo2}\label{zmoins}
\ee
\subsection{Mean field results}
\begin{figure}
\begin{center}
\includegraphics*[width=.65\textwidth]{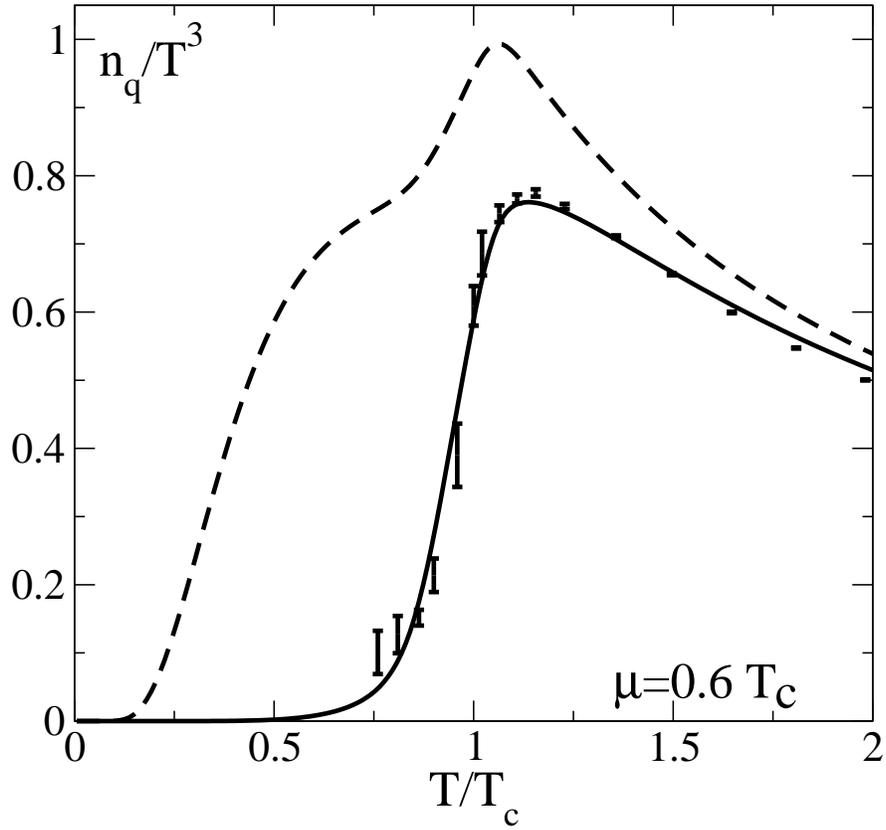}
\caption{
PNJL (solid line), NJL (dotted line) and lattice results (points)
for the net quark density at $\mu=0.6 T_c$ (from \cite{Ratti:2005jh}). 
\label{fig:QuarkDensity} 
}
\end{center}
\end{figure}

The solutions of the mean field equations are obtained by minimizing the grand potential
with respect to $m$, $\Phi$ and $\bar\Phi$, namely (again below $N_c=3$)
\be\label{eq:stationary}
\frac{\partial\Omega}{\partial\Phi}&=& 0 \nonumber\\
&=&
	\frac{T^4}{2} (-b_2(T) \bar\Phi - b_3 \Phi^2 + b_4 \Phi \bar\Phi^2) \nonumber \\
&&- 6 N_f T \int_\Lambda\frac{\mathrm{d}^3p}{\left(2\pi\right)^3} \left\{ \frac {e^{-2(E_p-\mu)/T}}{  1 + 3\left( \bar\Phi +
	  \Phi \expmT \right) 
\expmT + \expmmmT } \right. \nonumber \\
&& \qquad \qquad			
 \left. + \frac {e^{-(E_p+\mu)/T}}{  1 + 3\left( \Phi + \bar\Phi \exppT \right) \exppT + \exppppT }
		\right\} ,			
\label{eq:domegadfi}
\ee
\be
\frac{\partial\Omega}{\partial\bar\Phi}&=&0 \nonumber\\
&=&
	\frac{T^4}{2} (-b_2(T) \Phi - b_3 \bar\Phi^2 + b_4 \bar\Phi \Phi^2)  \nonumber \\
&&- 6 N_f T \int_\Lambda\frac{\mathrm{d}^3p}{\left(2\pi\right)^3} \left\{ \frac { e^{-(E_p-\mu)/T}}{  1 + 3\left( \bar\Phi +
	  \Phi \expmT \right)
 \expmT + \expmmmT } \right. \nonumber \\
&& \qquad \qquad		
\left. + \frac {e^{-2(E_p+\mu)/T}}{  1 + 3\left( \Phi + \bar\Phi \exppT \right) \exppT + \exppppT }
		\right\} \label{eq:domegadfibar}
\ee
and
\beq
\frac{\partial\Omega}{\partial m}&=& 0 
\label{eq:domegadm}
\eeq
which coincides with the gap equation (\ref{eq:Hartree_brut}).
A complete discussion of the results in mean field approximation is
given in 
\cite{Ratti:2005jh}.
For the purpose of this article, we only briefly discuss the result obtained in \cite{Ratti:2005jh} for the net quark number density,
defined by the equation
\begin{equation}
\frac{n_q\left(T,\mu\right)}{T^3}=-\frac{1}{T^3}\frac{\partial\Omega
\left(T,\mu\right)}{\partial\mu},
\end{equation}
that we display in Fig. \ref{fig:QuarkDensity}\footnote{Indeed in 
\cite{Ratti:2005jh} a different regularization procedure was employed with 
respect to the choice adopted in this paper. Namely, no cut-off was used for 
the finite $T$ contribution to the thermodynamical potential. This choice was 
made in order to better reproduce lattice results up to temperatures 
$T\sim 2T_c$. In any case, for lower temperature this difference in the 
regularization is unimportant. In particular our qualitative discussion of the 
role of the field $\Phi$ in mimicking confinement is independent of these 
details.}.
Note that an implicit $\mu$-dependence of $\Omega$ is also 
contained in the effective quark mass $m$ and in the expectation 
values $\Phi$ and $\bar\Phi$. Nevertheless, due to stationary equations 
(\ref{eq:domegadfi},~\ref{eq:domegadfibar},~\ref{eq:domegadm}), 
only the explicit dependence arising from the 
statistical factors has to be differentiated.

One can see that the NJL model (corresponding to the $\Phi\to 1$ limit
of PNJL) badly fails in reproducing the lattice findings, while the
PNJL results provide a good approximation for them.  One realizes
that, at a given value of $T$ and $\mu$, the NJL model always
overestimates the baryon density, even if, for large temperatures,
when in PNJL $\Phi\to1$, the two models merge.

On the other hand in the PNJL model below $T_c$ (when
$\Phi,\bar\Phi\to 0$) one can see from Eqs. (\ref{eq:termo1}) and
(\ref{eq:termo2}) that contributions coming from one and two
(anti-)quarks are frozen, due to their coupling with $\Phi$ and
$\bar\Phi$, while three (anti-)quark contributions are not suppressed
even below $T_c$.  This implies that, at fixed values of $T$ and
$\mu$, the PNJL value for $n_q$ results much lower than in the NJL
case. In fact all the possible contributions to the latter turn out to
be somehow suppressed: the one- and two-quark contributions because of
$\Phi,\bar\Phi\to 0$, while the thermal excitation of three quark
clusters has a negligible Boltzmann factor.

One would be tempted to identify these clusters of three dressed
(anti-)quarks with precursors of (anti-)baryons. Indeed no binding for
these structures is provided by the model. In any case it is
encouraging that coupling the NJL Lagrangian with the Polyakov loop
field leads to results pointing into the right direction.

In the following Section we explore the PNJL results in the mesonic
sector, investigating whether coupling the (anti-)quarks with the
$\Phi$ field constrains the dressed $q\bar{q}$ pairs to form stable
colorless structures.

\section{Mesonic correlators \label{correlators}}

In this Section, we address the central topic of our paper, i.e. the study of
correlators of current operators carrying the quantum numbers of
 physical mesons. We focus our attention on two particular cases:
the pseudoscalar iso-vector current
\be
  {J_P}^a(x) &=& \bar q(x) i \gamma_5 \tau^a q(x)\quad \mbox{ (pion)}
\ee
and the scalar iso-scalar current:
\be
  {J_S}(x) &=& \bar q(x) q(x) - \langle\bar q(x) q(x) \rangle 
\quad\mbox{ (sigma)} .
\ee
These are in fact the channels of interest to study the chiral
symmetry breaking-restoration pattern. In particular the scalar
current represents the fluctuations of the order parameter.

In terms of the above currents, the following mesonic correlation functions and
 their Fourier transforms are defined:
\beq
C^{PP}_{ab} (q^2) \equiv  i \int d^4x e^{i q.x}
 \exvalvide{\T \left( 
J^a_P(x) J^{b\dagger}_P(0) \right) }                 = C^{PP}(q^2) \delta_{ab}
\eeq
and
\beq
C^{SS}(q^2) \equiv  i \int d^4x e^{i q.x} \exvalvide{\T 
\left( J_S(x) J^\dagger_S(0) \right) }.
\eeq
In the above equations, the expectation value is taken with respect to the
 vacuum 
state and $T$ is the time-ordered product.

\subsection{Schwinger -- Dyson equations at $\boldsymbol{T=\mu=0}$}
Here we briefly summarize the usual NJL results for the mesonic correlators 
\cite{Oertel:2000jp,Klevansky:1997dk,Schulze:1994fy,Davidson:1995fq}, which we 
are going to generalize in Sec. (\ref{section:finiteT}) by including the case 
in which quarks propagate in the temporal background 
gauge field related to the Polyakov loop. 
The Schwinger -- Dyson equation for the meson correlator $C^{MM}$ is solved in 
the ring approximation (RPA):
\be
  C^{MM}(q^2) &=& \Pi^{MM}(q^2) + \sum_{M'} \Pi^{MM'} (2G_1) C^{M'M}
\ee
where the
\be
  \Pi^{MM'} & \equiv & \idp \Tr \left( \Gamma_M S(p+q) \Gamma_{M'} S(q) \right) 
\ee
are the one loop polarizations and $S(p)$ is the Hartree quark
propagator.  In terms of diagrams, one defines:
\be
  \Pi^{MM'} &=& \Gamma_M \diag{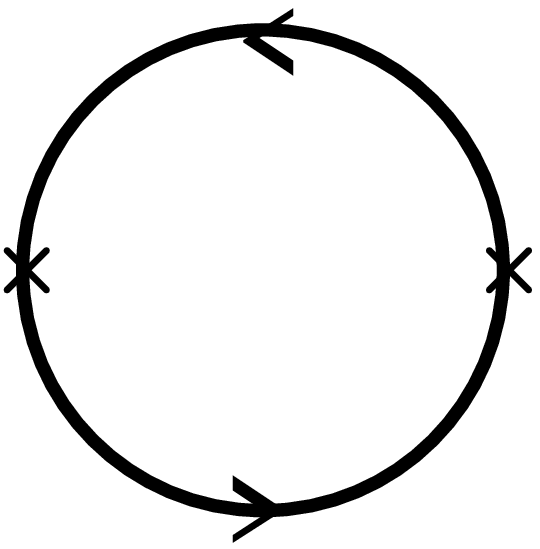} \Gamma_{M'}
\ee
and
\be
   C^{MM} &=& \diagscale{2cm}{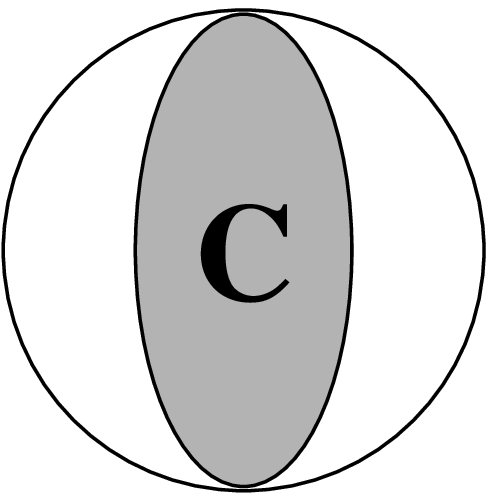} = \diagscale{2cm}{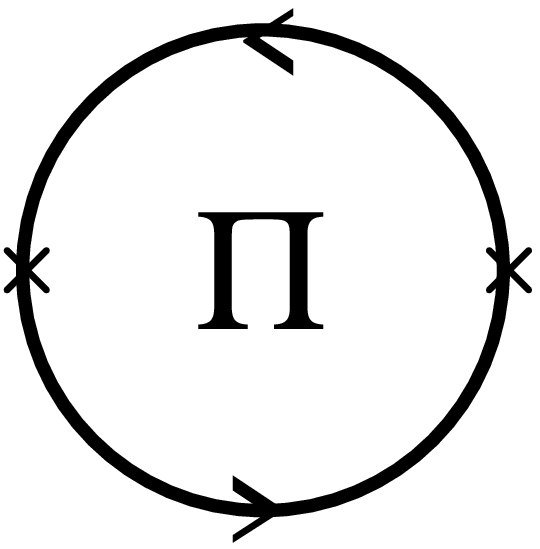} +  \diagscale{4cm}{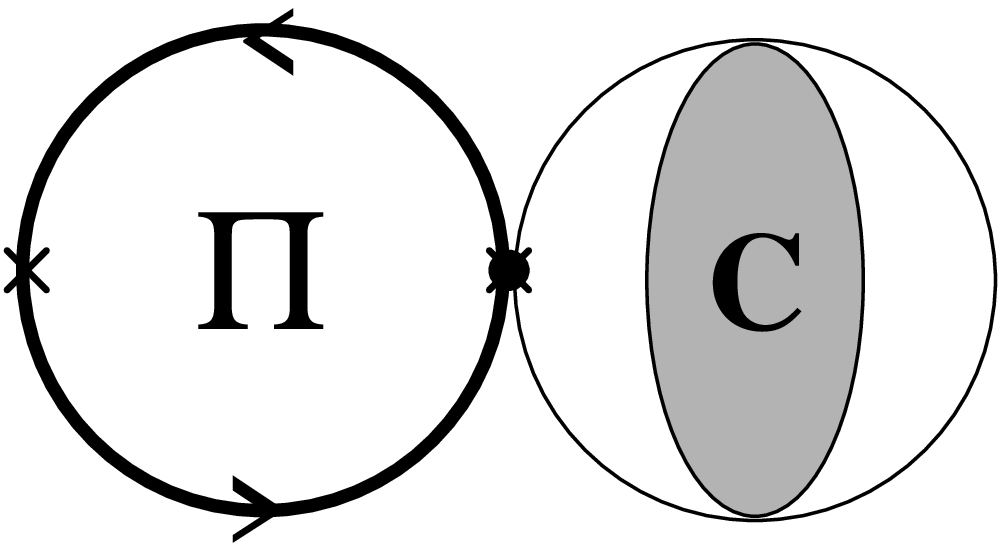} .
\ee

Hence, we need the following (one loop) polarization functions:
\be
\Pi^{PP}_{ab} (q^2) &=& \idp \Tr \left( i \gamma_5 \tau^a S(p+q) i \gamma_5 \tau^b  S(q) \right) = \Pi^{PP} (q^2) \delta_{ab} \\
\Pi^{SS} (q^2) &=& \idp \Tr \left( S(p+q) S(q) \right) .
\ee
Thus, for example, for the pion channel:
\be
\Pi^{PP} (q^2) &=& - 4 i N_c N_f \idp \frac{m^2 - p^2 + q^2/4}{[(p+q/2)^2 - m^2][(p-q/2)^2 - m^2]} \\
               &=& 4 iN_c N_f I_1 - 2iN_c N_f q^2 I_2(q^2)
\nonumber
\ee
the loop integrals being:
\be
  I_1 &=& \idp \frac{1}{p^2-m^2} \label{i1} \\
  I_2(q^2) &=& \idp \frac{1}{\left[(p+q)^2-m^2\right]\left[p^2-m^2\right]}  \label{i2} .
\ee
By defining\footnote{$\, \fsqp(q^2=0)$ is the pion decay constant
  $f^2_\pi$ in the chiral limit \cite{Klevansky:1992qe}.}:
\be
  \fsqp(q^2) &=& -4i N_c m^2 I_2(q^2) 
\ee
and owing to the fact that the Hartree equation (\ref{eq:Hartree}) implies
\be
  I_1 &=& \frac{m - m_0}{8i G_1 m N_c N_f},
\ee
one shows that \cite{Klevansky:1997dk}
\be
 \Pi^{PP}(q^2) &=& \frac{m-m_0}{2G_1m} + \fsqp(q^2) \frac{q^2}{m^2} \\
 \Pi^{SS}(q^2) &=& \frac{m-m_0}{2G_1m} + \fsqp(q^2) \frac{q^2-4m^2}{m^2} .
\ee
The explicit solutions of the Schwinger--Dyson equations in ring approximation
then read:
\bi
 \item{Scalar iso-scalar sector}
 \be
   C^{SS}(q^2) &=& \Pi^{SS}(q^2) + \Pi^{SS}(q^2) (2 G_1) C^{SS}(q^2) \\
   \Rightarrow C^{SS} &=& \frac{\Pi^{SS}(q^2)}{1 - 2G_1 \Pi^{SS}(q^2)} .
 \ee
 \item{Pseudo-scalar iso-vector sector }
 \be
   C^{PP}(q^2) &=& \Pi^{PP}(q^2) + \Pi^{PP}(q^2) (2 G_1) C^{PP}(q^2) \\
   \Rightarrow C^{PP} &=& \frac{\Pi^{PP}(q^2)}{1 - 2G_1 \Pi^{PP}(q^2)} .
 \ee
\ei

\subsection{NJL Schwinger-Dyson equations at finite $T$ and $\mu$}

In order to study the problem at finite temperature and baryon density
in the imaginary time formalism ($t=-i\tau$ with $\tau\in[0,\beta]$), the 
$\tau$-ordered
product of the operators replaces the usual time-ordering and all the 
expectation values are taken over the grand-canonical ensemble.

One can decompose all the integrands, for example in $I_2$, as a sum of 
partial fractions of the form 
\be
 \frac{1}{i\omega_n -E + \mu}.
\label{eq:frac}
\ee
The sum over Matsubara frequencies is then computed by using:
\be
 \frac1{\beta} \sum_n  \frac{1}{i\omega_n - E + \mu} = f(E - \mu) 
\ee
where the Fermi -- Dirac distribution function is given by:
\be
  f(E) = \frac{1}{1+e^{\beta E}} .
\ee

The integrals $I_1$ and $I_2$ (Eqs. \refp{i1} and \refp{i2}) at finite temperature and density 
are then expressed as 
\cite{Oertel:2000jp,Hatsuda:1986gu,Florkowski:1993br,Su:1990dy,Farias:2004tf}:
\be
  I_1 &=& -i \idpT \frac{\dtherm}{2E_p} \\
  I_2(\omega, \vec q) &=& i \idpT 
	\frac{1}{2E_p 2E_{p+q}} \frac{\fp p + \fm p - \fp{p+q} - \fm{p+q}}{\omega - E_{p+q} + E_p} \nonumber \\
       && \hspace*{-1truecm}+ i \idpT \frac{1 - \fm p - \fp{p+q}}{2E_p 2E_{p+q}}
	\left(
	\frac{1}{\omega + E_{p+q} + E_p} - \frac{1}{\omega - E_{p+q} - E_p} \right)
\ee
(these expression are implicitly taken at $\omega \rightarrow \omega+i
\eta$ to obtain retarded correlation functions).

Then all the zero temperature results can be continued to finite
temperature and density by a redefinition of $I_1$ and $I_2$.

At $\vec q = \vec{0}$, the integral $I_2$ reduces to:
\beq
I_2\left(\omega,\vec{0}\right)=
-i\idpT\frac{1-\fp p-\fm p}{E_p\left(\omega^2-4E_{p}^{2}\right)}
\eeq
so that we obtain:
\beq
\Pi^{PP}\left(\omega,\vec{0}\right)&=&-8N_cN_f\idpT\frac{E_p}{\omega^2-
4E_{p}^{2}}\left(1-\fp p-\fm p\right) \\
\Pi^{SS}\left(\omega,\vec{0}\right)&=&-8N_cN_f\idpT\frac{1}{E_p}
\frac{E_p^2-m^2}{\omega^2-
4E_{p}^{2}}\left(1-\fp p-\fm p\right) .
\eeq
It then follows:
\beq
  \Imag \left(-i I_2(\omega, 0) \right) &=&
    \frac{1}{16\pi} \left(1 - f\left(\frac \omega 2 - \mu\right) 
                            - f\left(\frac \omega 2+\mu\right) \right) 
    \sqrt{\frac{\omega^2-4m^2}{\omega^2}} \nonumber \\
  && \times \Theta(\omega^2-4m^2)\Theta(4(\Lambda^2+m^2)- \omega^2)
\label{ImI1}
\eeq
(and of course, the real part is given by the Cauchy principal value of the integral).
Hence:
\be
 &&\Imag \Pi^{PP}(\omega, 0) = 2 N_f N_c \omega^2 \Imag (-i I_2(\omega)) 
\nonumber \\
    &&\qquad  = \frac{N_c N_f \omega^2}{8\pi} 
    \sqrt{\frac{\omega^2-4m^2}{\omega^2}}\, N(\omega,\mu)\,
 \Theta(\omega^2-4m^2)\Theta(4(\Lambda^2+m^2)- \omega^2)
\label{eq:ImPP} \\
&&\Imag \Pi^{SS}(\omega, 0) =\nonumber \\
   && = \frac{N_c N_f (\omega^2-4m^2)}{8\pi} 
    \sqrt{\frac{\omega^2-4m^2}{\omega^2}}\,  N(\omega,\mu)\, \nonumber 
\Theta(\omega^2-4m^2)\Theta(4(\Lambda^2+m^2)- \omega^2)\\
\label{eq:ImSS} 
\ee
with 
\be
  N(\omega, \mu) = \left(1 - f\left(\frac \omega 2 - \mu\right) 
                            - f\left(\frac \omega 2+\mu\right) \right) .
\ee

\subsection{PNJL Schwinger-Dyson equations at finite $\boldsymbol T$ 
and $\boldsymbol \mu$ \label{section:finiteT}}

Here we derived explicitly the expressions for the modified Fermi--Dirac distribution
functions Eqs.(\ref{fpPhi}) and (\ref{fmPhi}).

Again, all the summation over Matsubara frequencies can be reduced to
the sum of fractions like \refp{eq:frac}. By defining:
\be
  F(E_p - \mu + i A_4) &\equiv &\frac1{\beta}\sum_n \frac{1}{i\omega_n -E_p+\mu-i A_4}
\ee
one shows that:
\beq
&& \Tr_c F(E_p - \mu + i A_4) \nonumber \\ 
&& \qquad = f(E_p - \mu +i (A_{4})_{11}) + 
f(E_p - \mu +i (A_{4})_{22}) + f(E_p - \mu +i (A_{4})_{33})
\eeq
where $(A_4)_{ii}$ are the elements of the diagonalized
$A_4$ matrix.

Let us write the Fermi--Dirac distribution function according to:
\be
  \fm p &\equiv & -\frac1{\beta} \partialD{z^+}{E_p}\;,
\ee
where 
\be
z^+ &\equiv & \ln\left(1+\expm\right)
\ee
can be viewed as a density of partition function.
We then obtain
\be
 \Tr_c F(E_p - \mu + i A_4) &=& -\frac1{\beta} \sum_i \partialD{\ln\left(1+\expm 
e^{- i \beta (A_4)_{ii}}\right)}{E_p} = -\frac1{\beta}\, \Tr_c \partialD{\ln\left(1+\expm e^{- i \beta A_4} \right)}{E_p} 
\nonumber\\
        &=& -\frac1{\beta}\, \Tr_c \partialD{\ln\left(1+L^\dagger \expm\right)}{E_p} = -\frac1{\beta}\, \partialD{z_\Phi^+}{E_p} \label{eq:TrF}
\ee
where $z_\Phi^+ = \ln\left(1+L^\dagger \expm\right)$ is the corresponding density of partition function in PNJL (already
introduced in Eq.\refp{zplus}).
Hence 
\be  
\Tr_c F(E_p - \mu + i A_4) &=&
  3 \frac{ \left( \bar\Phi + 2\Phi \expm \right) \expm + \expmmm }
  {1 + 3\left( \bar\Phi + \Phi \expm \right) \expm + \expmmm} .
\ee
We can do the same for the $F(E_p + \mu - i A_4)$ case. \\

Hence, we can define:
\be
  f^+_\Phi(E_p) &\equiv& \frac{1}{N_c}  \Tr_c F(E_p - \mu + i A_4) = 
- \frac{1}{\beta N_c} \partialD{z_\Phi^+}{E_p} 
\ee
and 
\be 
  f^-_\Phi(E_p) &\equiv& \frac{1}{N_c}  \Tr_c F(E_p + \mu - i A_4) = - \frac{1}{\beta N_c} \partialD{z_\Phi^-}{E_p}\;, 
\ee
where $z_\Phi^+$ and $z_\Phi^-$ are the densities \refp{zplus} and \refp{zmoins} of the partition function in PNJL.

The only changes in going from NJL to PNJL can then be summarized in the 
following prescriptions:
\be
  f(E_p-\mu) &\Longrightarrow& f^+_\Phi(E_p) 
          =\frac{ \left( \bar\Phi + 2\Phi \expm \right) \expm + \expmmm }
  {1 + 3\left( \bar\Phi + \Phi \expm \right) \expm + \expmmm} 
\label{f} \\
  f(E_p+\mu) &\Longrightarrow& f^-_\Phi(E_p) 
          =\frac{ \left( \Phi + 2\bar\Phi \expp \right) \expp + \expppp }
  {1 + 3\left( \Phi + \bar\Phi \expp \right) \expp + \expppp} .
\label{fbar}
\ee
Of course in the above the corresponding PNJL quark mass $m$, given
by the Hartree equation with these modified distribution functions, should be used.

\begin{center}
\begin{figure}
\includegraphics[width=.85\textwidth]{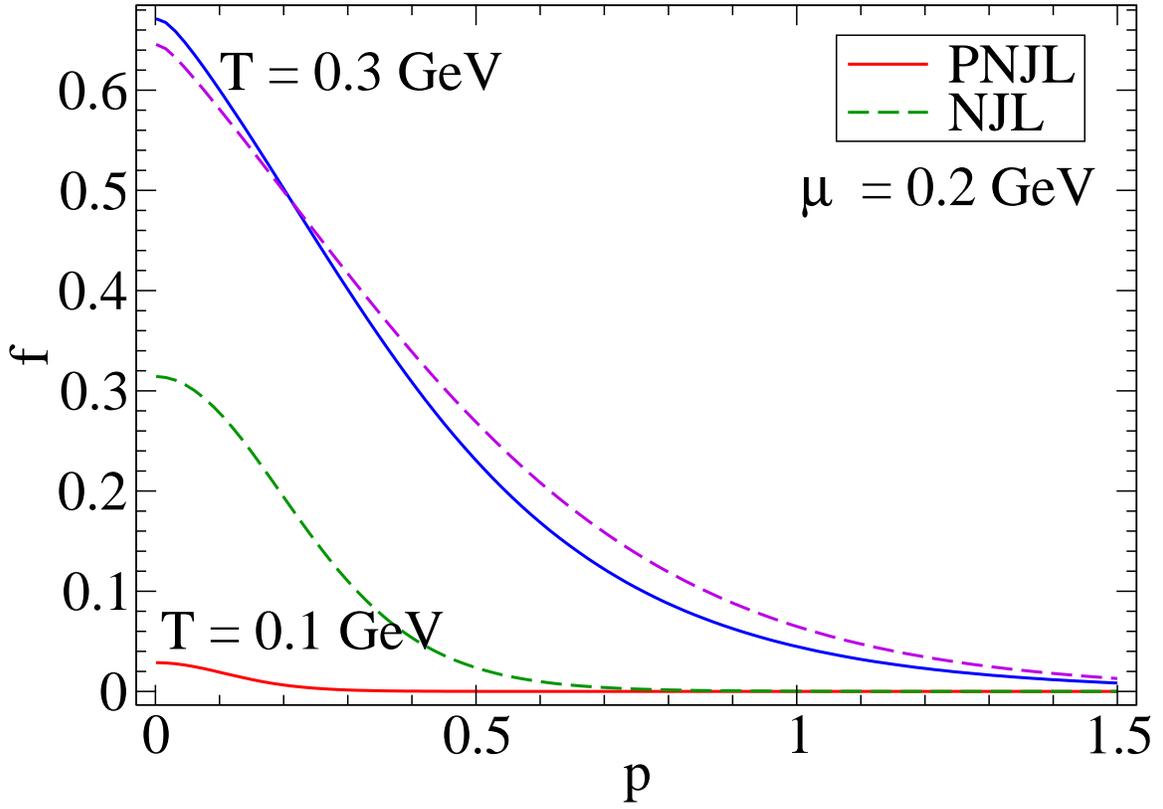} \null \\
\vspace*{0.2truecm}
\caption{Fermi -- Dirac distribution function $f(E_p-\mu)$ (valid for
the NJL model) and the corresponding function $f^+_\Phi(E_p)$ (valid
for the PNJL one) as functions of $p$, for different
temperatures.  $\Phi$, $\bar\Phi$ and m are taken at their mean field
values. The upper lines refer to $T=0.3$ GeV, the lower ones to $T=0.1$ GeV.}
\label{fig:fPhi}
\end{figure}
\end{center}

The functions $f(E_p-\mu)$ and $f^+_\Phi(E_p)$ are displayed in
Fig.~\ref{fig:fPhi} for two different temperatures versus $p$, keeping
$\Phi$, $\bar\Phi$ and $m$ at the mean field values. For temperatures
smaller than $T_c$ (for example $T=0.1$ GeV $\simeq T_c/2$), the effect
of the Polyakov loop turns out to be more relevant than for larger
temperatures, close to $T_c$.

In discussing the PNJL results for the net quark density we already
stressed the role of $\Phi$ and $\bar\Phi$ in suppressing one and two
(anti-)quark clusters in the confined phase. This also emerges in
Fig.~\ref{fig:fPhi} comparing the PNJL and NJL curves at
$T=0.1$~GeV. Clearly the two models differ substantially when $\Phi
,\bar\Phi\to 0$. On the contrary, as $\Phi ,\bar\Phi\to 1$ they lead
to similar results.

We conclude this Section by stressing once more that the recipes given in
Eqs.~\refp{f} and \refp{fbar} allow one to straightforwardly
generalize NJL results to the PNJL case.

\section{Meson spectral function and propagator \label{spectral}}

In the rest of the paper we present our numerical results for the
masses and spectral functions of the scalar ($\sigma$) and
pseudoscalar ($\pi$) mesons in a hot and dense environment.

The spectral (or strength) function $F^{MM}$ of the correlator
$C^{MM}$ is defined according to:
\be
 F^{MM}(\omega, \vec q) \equiv \Imag C^{MM}(\omega+i\eta,\vec{q}) =
 \Imag \frac {\Pi^{MM}(\omega+i\eta,\vec{q})} {1- 2 G_1
 \Pi^{MM}(\omega+i\eta,\vec{q})} .
\ee
For the sake of simplicity, in the following we will consider only the
zero momentum case: hence we will drop the dependence on $\vec{q}$.
One gets:
\beq
F^{MM}(\omega)=\frac{\pi}{2G_1}\frac{1}{\pi}\frac{2G_1\Imag \Pi^{MM}(\omega+i\eta)}
{\left(1-2G_1 \Real\Pi^{MM}(\omega)\right)^2+\left(2G_1\Imag \Pi^{MM}(\omega+i\eta)\right)^2} .
\eeq
For $\omega<2m(T,\mu)$, $\Imag\Pi=0$ hence the decay channel into a dressed $q\bar q$ pair
is closed and the spectral function gets a bound state
contribution expressed by a delta peak in correspondence of the mass
of the meson.
Indeed:
\beq
F^{MM}(\omega) = \frac{\pi}{2G_1}\delta\left(1-2G_1 \Real{\Pi^{MM}}(\omega)\right) =
\frac{\pi}{4G_1^2{
\left|\frac{\partial\Real{\Pi^{MM}}}{\partial\omega}\right| }_{\omega=
m_M}}\delta(\omega-m_M)\; .
\eeq
and the meson mass $m_M$ is the solution of the equation
\beq
1 - 2 G_1 \Real \Pi^{MM}(m_M)=0\;\label{eq:mesonmass} .
\eeq
On the other hand, for $\omega>2m(T,\mu)$, $\Imag \Pi\ne 0$ and the meson
spectral function gets a continuum contribution. Thus if the
solution of Eq. (\ref{eq:mesonmass}) occurs above such a threshold,
then the spectral function will still present a peak characterized by
a width related to the decay channel $M\to q\bar q$. In such a case
the meson is no longer a bound, but simply a resonant state.
If $\Imag \Pi$ stays almost constant around the position
of the peak, the spectral function is well approximated by a Lorentzian
with a width given by:
\beq
\Gamma_M=2G_1\Imag \Pi^{MM}(m_M)\;.
\eeq 

On the other hand, if $\Imag \Pi$ varies with $\omega$ the solution of
Equation (\ref{eq:mesonmass}) and the maximum of the spectral function
no longer coincide, the latter being typically below the former. 
 In the following we choose to identify the mass
of the meson with the maximum of the spectral function.

We notice that $C^{MM}$ is a correlator of current operators which is
the quantity investigated in lattice calculations. But one can also
get useful information concerning $q \bar q$ scattering processes by
extracting the meson propagator from the $T -$matrix
\cite{Oertel:2000cw}.

In the present framework it can be shown that the propagator for a meson is
\be
D_M(\omega) & = & - G_1 \frac{C^{MM}(\omega)}{\Pi^{MM}(\omega)}.
\ee
In the quasi-particle approximation, the above simplifies to:
\be
D_M(\omega) &\simeq& \frac{-i g^2_{M qq}}{\omega^2 - m^2_M}
\ee
where $m^2_M$ verifies the pole equation (\ref{eq:mesonmass})
and the effective meson$-$quark coupling constant,
\be
  g^{-2}_{M qq} &=& { \left. \partialD{\Pi^{MM}}{\omega^2} \right| }_{\omega = m_M} ,
\ee
is the residue at the pole.

\section{Numerical results \label{numerical}}

 In this Section we present, in the PNJL model, our numerical results
for the properties of the $\sigma$ and $\pi$ mesons in a hot and dense
environment.

The special role played by the $\sigma$ spectral function, embodying
the correlations among the fluctuations of the order parameter (the
chiral condensate), was first pointed out in~\cite{Hatsuda:1985eb},
within the NJL model.  In particular, it was shown that in the Wigner
(ordered) phase of chiral symmetry the $\sigma$ spectral function
(which becomes approximatively degenerate with the $\pi$ one, due to
chiral symmetry restoration) displays a pronounced peak, moving to
lower frequencies and getting narrower as $T\to T_c$ from above.  The
above excitations, characterizing the regime of temperatures slightly
exceeding $T_c$, were then identified as \emph{soft modes},
representing a precursor phenomenon of the phase transition.

We will show in the following that the above qualitative features of
the mesonic excitations are preserved, once the coupling with the
Polyakov loop field is introduced in the NJL model.

\subsection{NJL vs. PNJL: Characteristic temperatures\label{NJLvsPNJLtemperatures}}
Before discussing the mesonic properties, we need to identify the
characteristic temperatures which separate the different
thermodynamic phases in PNJL and NJL. In order to define a
``critical'' temperature one would like to refer to the order
parameters (vanishing in the disordered-symmetric phase and
non-vanishing in the ordered-broken phase). The latter, as already
pointed out, can be identified with the Polyakov loop $\Phi$ (if
$m_q\to\infty$) and with the chiral condensate $\langle\bar{q}q\rangle$
(if $m_q\to 0$), for the deconfinement and chiral phase transitions,
respectively. Chiral symmetry restoration is also signalled by $m_\pi
= m_\sigma$ (or, strictly speaking, by the merging of $\pi$ and
$\sigma$ spectral functions).

In the present context, chiral symmetry is explicitly broken by the
presence of a finite bare quark mass; 
nevertheless the discontinuity displayed by the order parameter or
by its derivatives still allows one to define a critical line
in the $(\mu - T)$ plane separating the two phases:
for small values of the baryo-chemical potential the transition
is known to be a continuous one (cross-over) but for larger
densities it becomes first order.
A critical temperature, identified with the
maximum of $- d m / d T$ ($\Tcc$), is then commonly used in the 
literature: this can be
applied both to NJL and PNJL.  The PNJL model also displays a similar
cross-over for the effective Polyakov loop $\Phi$ (which was shown to
occur at a temperature close to $\Tcc$ in
Ref.~\cite{Ratti:2005jh}). We also notice that, in the low-density
limit, two-flavour lattice QCD shows a cross-over, occurring at the same
\emph{critical} temperature both for the
deconfinement and the chiral transitions.

A different choice for a common \emph{characteristic} temperature is also
possible, corresponding to the minimum of $m_\sigma(T)$ (typical
temperature where the pion and sigma spectral functions start to merge); we 
denote it as $\Tms$. 

In order to compare the NJL and PNJL results, it is useful to follow
the evolution of the observables as functions of the temperature,
expressed both in physical units (MeV) as well as rescaled in units of
a {characteristic} temperature.  For the latter we choose the
corresponding $\Tms$ in NJL and PNJL. Our choice is motivated in
Sec.~\ref{widths}.

These temperatures computed in PNJL and NJL (at $\mu = 0$), together
with the Mott temperatures for the pion (the temperature at which
the decay of a pion into a $\bar{q} q$ pair becomes energetically
favourable), are quoted in Table
\ref{table:temperatures}. We remind that in the framework of PNJL a
further critical temperature related to the deconfinement phase
transition can be defined. Its value $T_c^\Phi = 0.250$ GeV
corresponds to the $\Phi$ crossover location.
Worth noticing is that the value of the latter differs by only $6$ MeV from
$\Tcc$. 
As it is evident from Table~\ref{table:temperatures}, the
characteristic temperatures that we get in the present approach are
much larger than the one which is usually quoted for the
chiral/deconfinement phase transition in two flavour QCD ($T_c\simeq
173$ MeV) on the basis of lattice calculations. They are also larger
than the ones obtained in the PNJL model in Ref.~\cite{Ratti:2005jh}.
This difference is due to our choice to regularize both the zero
\emph{and} the finite temperature contributions with a
three-dimensional momentum cutoff, and also to the fact that we do not
rescale the parameter $T_0$ to a smaller value when we introduce quarks in the
system. Nevertheless, for the purpose of the present work, the
absolute value of the critical temperature is not important: the
general properties of mesons that we discuss here are in fact
independent of the specific value of $T_c$.

\begin{table}
\begin{center}
\begin{tabular}{|c|c|c|c|c|}
\hline
PNJL & $\Tcc = 0.256$ GeV & $\Tms = 0.277$ GeV & $T_{Mott} = 0.272 $ GeV $= 1.06 \Tcc$ \\
\hline
 NJL & $\Tcc = 0.194$ GeV & $\Tms = 0.210$ GeV & $T_{Mott} = 0.212 $ GeV $= 1.09 \Tcc$ \\
\hline
\end{tabular}
\caption{Characteristic temperatures in the NJL and PNJL models at zero chemical potential.}
\label{table:temperatures}
\end{center}
\end{table}

\subsection{Mesonic masses and spectral functions}

\begin{center}
\begin{figure}
\doublefig
{\includegraphics[width=1.09\textwidth]{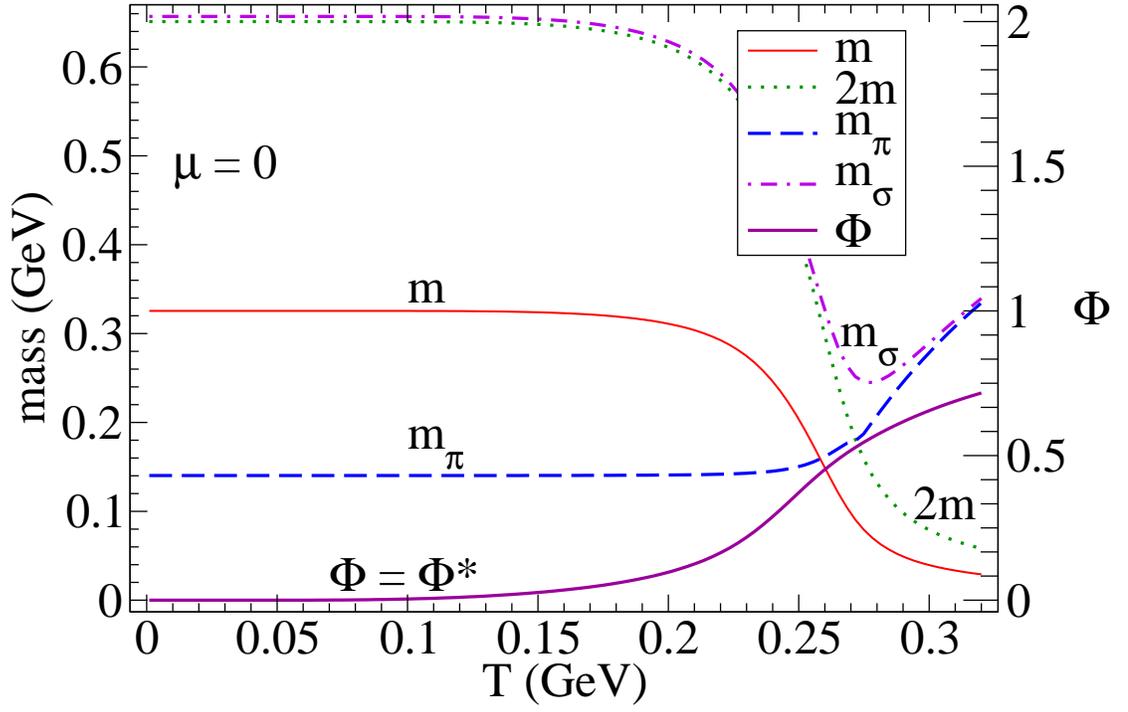}}
{\includegraphics[width=0.99\textwidth]{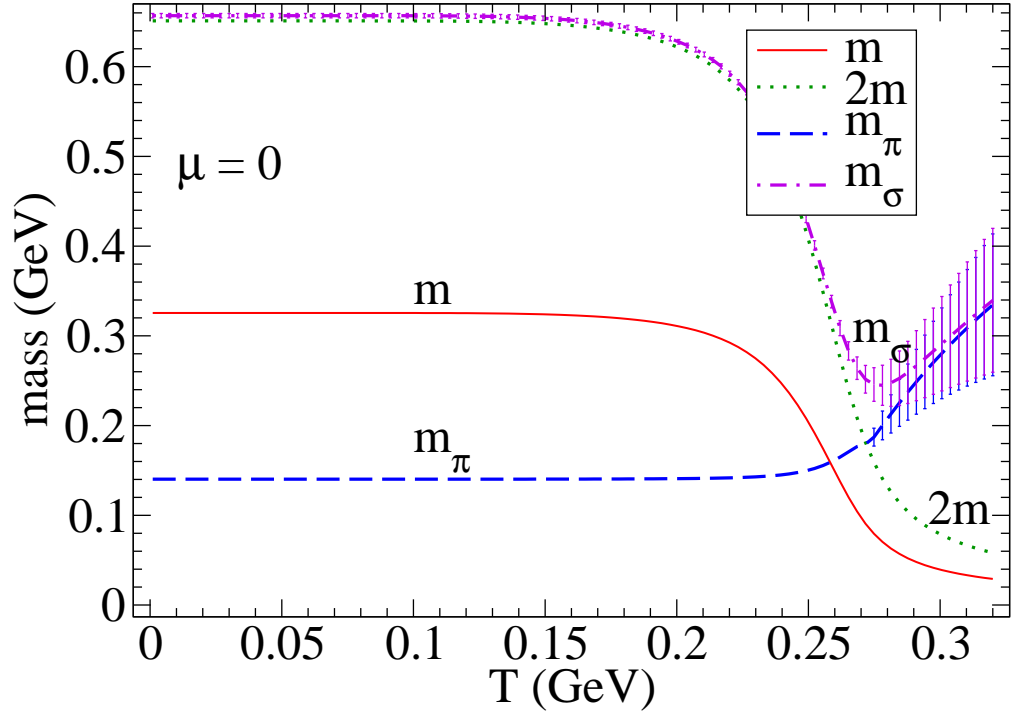}}
\caption{Top: Masses of the $\sigma$ and $\pi$ as functions of the
temperature, together with the Hartree quark mass and the Polyakov
loop, in the PNJL model at $\mu=0$.
The threshold $2m$ is also plotted to show that the $\sigma$
mass is close to this value below $0.25$ GeV.  Bottom:
Same as before, but without the Polyakov loop and adding instead the width
of the mesons, represented by error bars.}
\label{fig:MesonMasses1}
\end{figure}
\end{center}

\begin{figure}
\doublefig
{\includegraphics[width=0.95\textwidth]{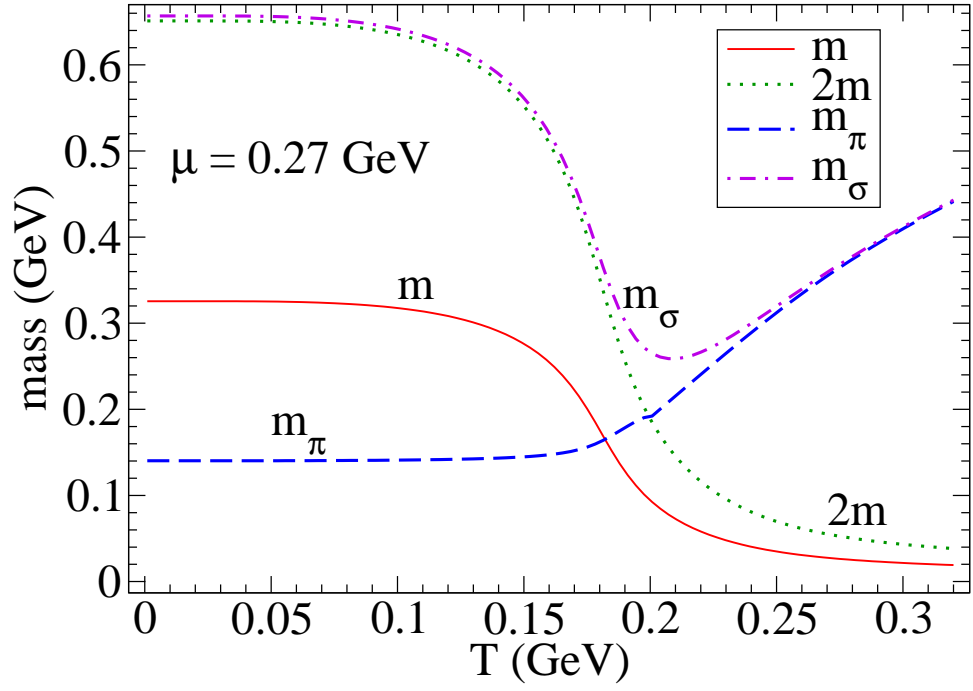}}
{\includegraphics[width=0.95\textwidth]{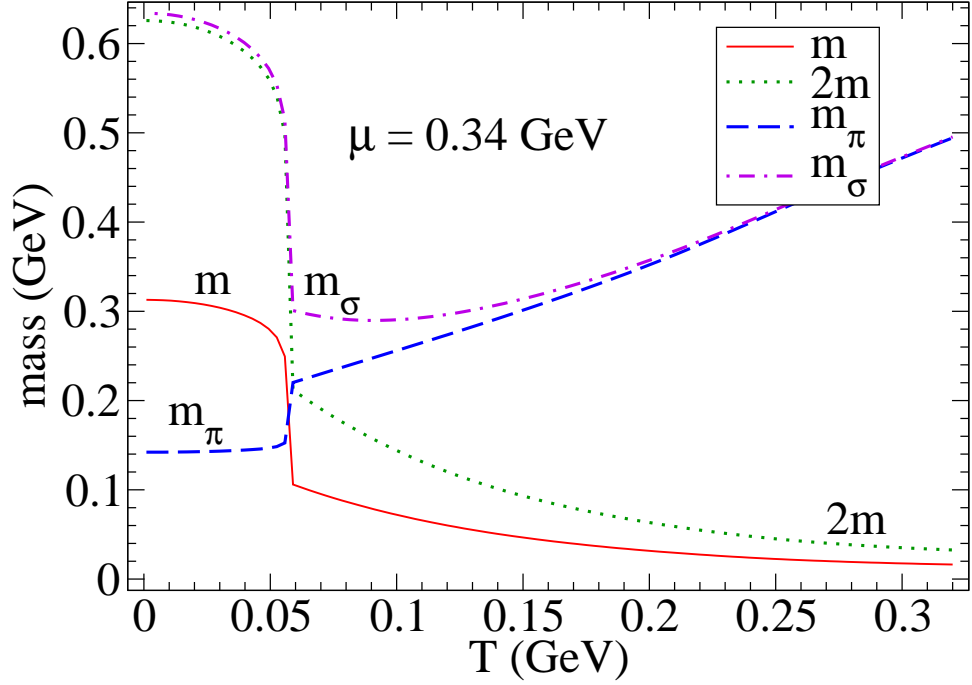}}
\caption{Masses of the $\sigma$ and $\pi$ as functions of the temperature, 
together with the Hartree quark mass,
in the PNJL model at $\mu=0.27$ GeV (top)
and $\mu = 0.34$ GeV (bottom).  The lower figure clearly displays a
first order phase transition related to the discontinuity of the
chiral condensate occurring at $\Tcc\simeq 0.06$~GeV. }
\label{fig:MesonMasses2}
\end{figure}

In Figs.~\ref{fig:MesonMasses1} and \ref{fig:MesonMasses2}, we plot the
masses of the $\sigma$ and $\pi$ mesons, together with the Hartree
quark mass and the Polyakov loop as functions of the temperature.  
The first evidence emerging from these 
figures is that the behavior of mesons in PNJL looks very similar to 
the corresponding one in NJL \cite{Hatsuda:1985eb,Hatsuda:1986gu,Costa:2003uu,
Costa:2002gk,Dorokhov:1997rv} 
(as it can be seen in Fig.~\ref{fig:NJLvsPNJLmass} 
where NJL and PNJL results are directly compared).

In Fig.~\ref{fig:MesonMasses1}, at $\mu=0$, the $\sigma-$mass closely
follows $2m$ below $T\simeq0.25$ GeV. Then the two curves decouple:
the mass of the dressed quarks approaches its current value, while the
mass of the $\sigma$ meson starts increasing.
  
The $\pi-$mass is small (it is a Goldstone boson if $m_0 = 0$) and
approximately constant at low temperature. Then it starts to increase
and tends to join the $\sigma-$mass above $T\simeq 0.25$ GeV.  Both
$\pi$ and $\sigma$ decay into $q\bar q$ as soon as $m_M > 2 m$. This
feature can be seen clearly in the lower panel of
Fig.~\ref{fig:MesonMasses1}, where the width of the mesons is shown,
together with their mass. Also at low temperatures, at  variance with a
realistic physical situation, the production of free $q \bar q$ pairs is
allowed due to the non-vanishing width of the $\sigma$ meson.

The two panels of Fig.~\ref{fig:MesonMasses2} show the behavior of the 
mesonic masses 
as functions of temperature, for two different values of the chemical 
potential.
For $\mu=0.27$ GeV the system undergoes a crossover from the low-temperature, 
chirally broken phase, to the high-temperature, chirally restored one, in 
analogy to what happens at $\mu=0$. As a consequence, the behaviour of the
mesonic masses is very similar to the one shown at vanishing chemical 
potential (see Fig.~\ref{fig:MesonMasses1}), the only difference being
a lower critical temperature as $\mu$ is increased.

The pattern changes, instead, at $\mu = 0.34$ GeV, where a
discontinuity in the masses {(reflecting an analogous behavior of the
chiral condensate $\langle\bar{q}q\rangle$)} appears. This can be
understood by observing that between $\mu = 0.27$ and $0.34$ GeV there
exists a critical point~\cite{Ratti:2006gh}, separating a crossover
from a first order phase transition.

\begin{figure}
\doublefig
{\includegraphics[width=\textwidth]{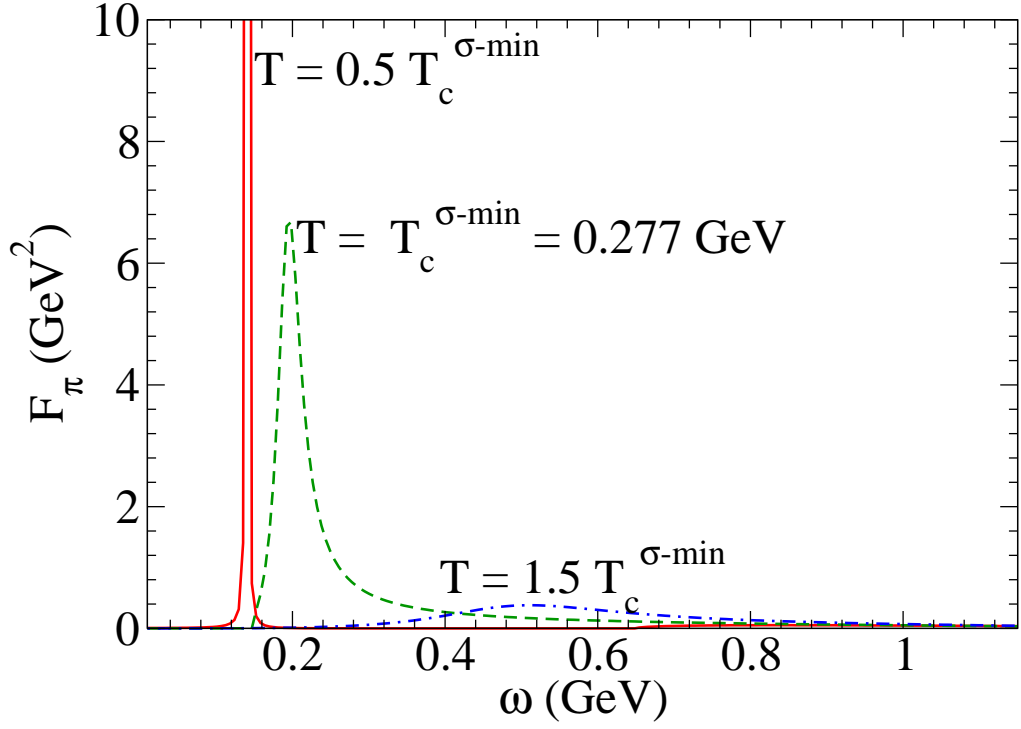}}
{\includegraphics[width=\textwidth]{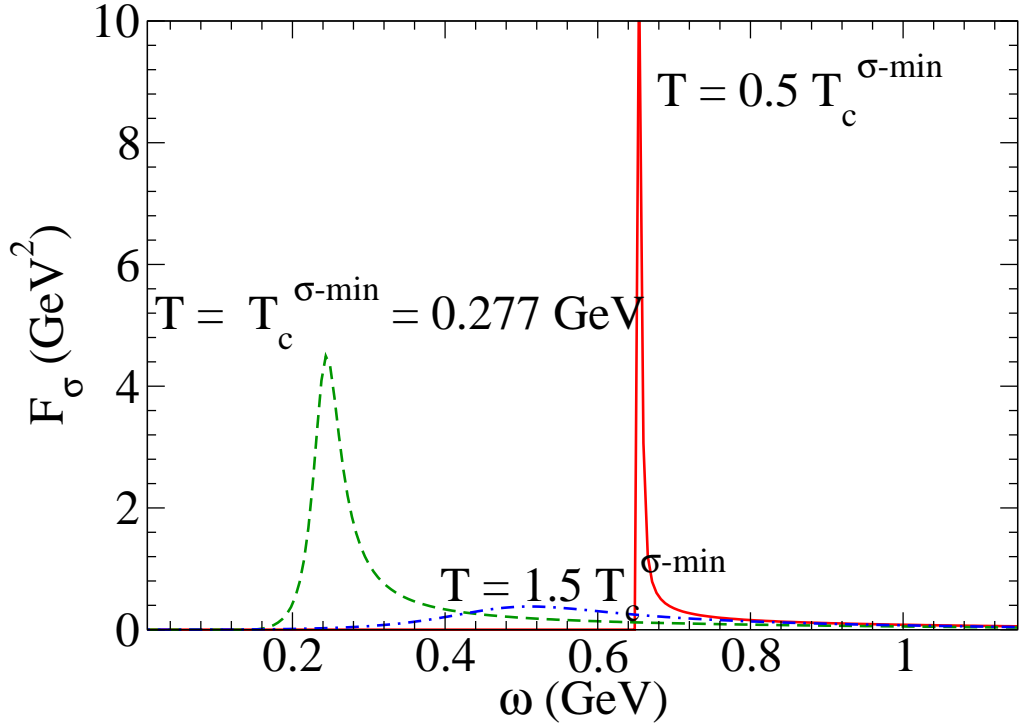}}
\caption{Spectral function $F^{MM}(\omega)$ of the pion (top) and
sigma (bottom) in PNJL, at $\vec q = \vec{0}$, as a function of $\omega$, for
different temperatures and $\mu = 0$. The width of the delta peak in
the $\pi$ spectral function when $m_\pi < 2m$ is artificially given by
a small, positive $i \eta$.}
\label{fig:SFs}
\end{figure}

In concluding this paragraph, we show the pion and 
$\sigma$ spectral functions in Fig.~\ref{fig:SFs}. Notice their progressive 
broadening as the 
temperature increases. Besides, they tend to
merge for $T >  \Tms$, in the chirally symmetric phase, as expected.

\subsection{NJL vs. PNJL: Mesonic masses at $\boldsymbol{\mu=0}$}

\begin{figure}
\bc
\includegraphics[width=0.8\textwidth]{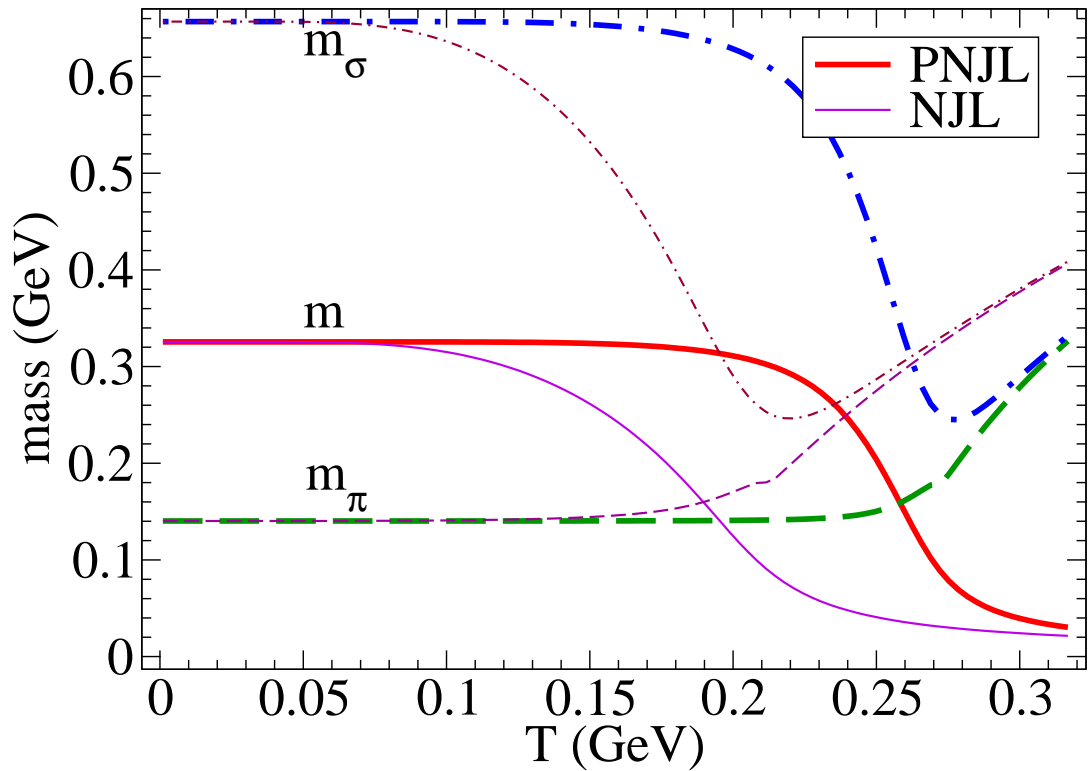}
\ec
\caption{$\sigma$ and
$\pi$ masses as functions of the temperature, together with the Hartree quark 
mass in NJL (thin lines) and PNJL 
(thick lines) model ($\mu=0$).}
\label{fig:NJLvsPNJLmass}
\end{figure}

In Fig.~\ref{fig:NJLvsPNJLmass} we show a direct comparison between
NJL and PNJL results for the mesonic masses at $\mu=0$.  According to
the features discussed in the previous paragraph, the key quantity
which governs the temperature evolution of the mesonic masses is the
dressed quark mass. As it is evident from the figure, the main
quantitative difference between the results of the two models is the
shift of the critical temperature for the phase transition, which
turns out to be higher in the PNJL model, with respect to the
``classic'' NJL one. From a qualitative point of view, there is a good
agreement between the results of the two models: in both cases in
fact, the $\sigma$ meson mass closely follows the behaviour of $2m$
for small temperatures, decreasing when one approaches the phase
transition region. Above $T_c$, instead, $m_\sigma$ increases with the
temperature, merging with the pion mass. This behavior reflects chiral
symmetry restoration, a feature which is correctly described by both
models, and therefore not spoiled by the coupling of quarks to the
Polyakov loop.

\subsection{NJL vs. PNJL: the  $\boldsymbol{\sigma}$ spectral 
function \label{widths}}

\begin{figure}

\doublefig
{\includegraphics[width=\textwidth]{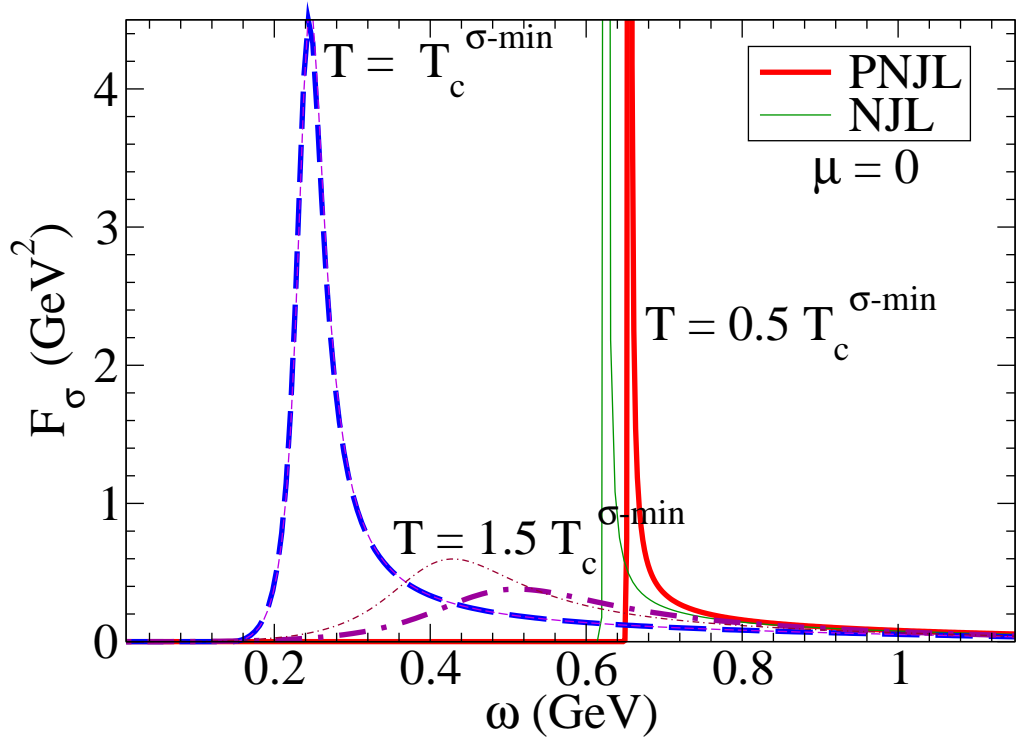}}
{\includegraphics[width=\textwidth]{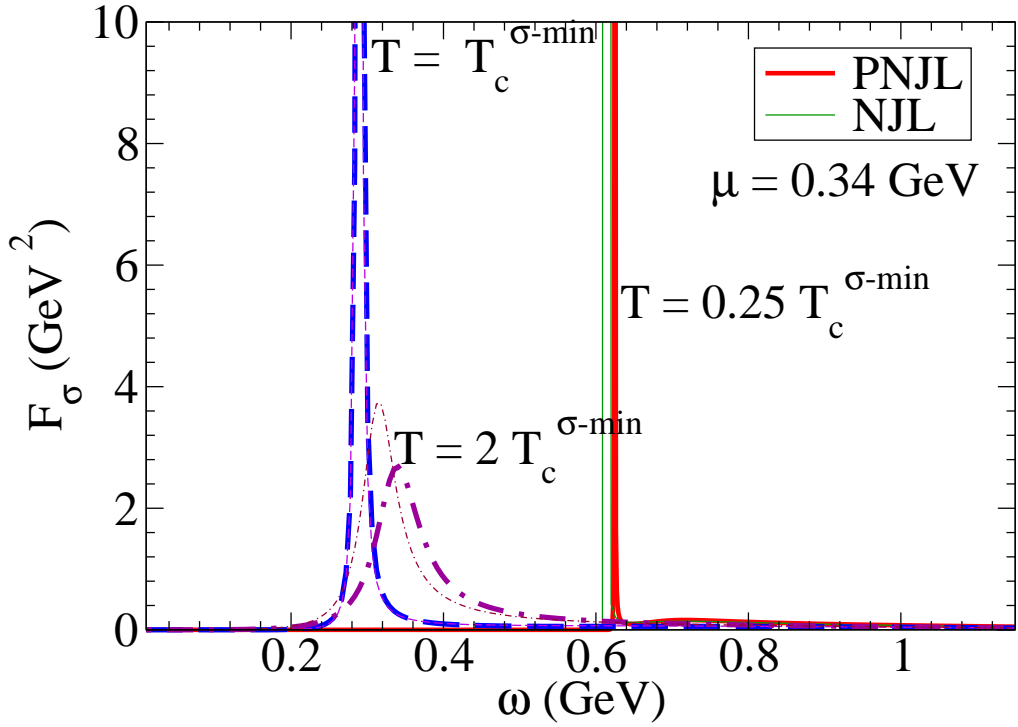}}
\caption{Comparison of the $\sigma$ spectral function at $\vec q = \vec{0}$, as
a function of $\omega$, for different temperatures and two chemical 
potentials. Thick lines correspond to PNJL model and thin ones to NJL.
Notice the almost perfect coincidence at $T=\Tms$.}
\label{fig:NJLvsPNJLsf}
\end{figure}

In this paragraph we discuss the width of the $\sigma$ spectral
function for the process $\sigma\rightarrow\bar{q}q$, as a function
of the reduced temperature $T/\Tms$. {As already anticipated, we find
it convenient to rescale $T$ by $\Tms$ since, interestingly enough,
the $\sigma$ spectral function computed in the PNJL model almost
coincides with the one evaluated in NJL at $T\simeq\Tms$.} This can be
clearly seen in Fig.~\ref{fig:NJLvsPNJLsf}, where the $\sigma$ spectral 
function
at $\vec{q}=\vec{0}$ is plotted vs. frequency. Notice the
broadening of the $\sigma$ spectral function in PNJL as compared to
the NJL one when $T>\Tms$, pointing to a stronger production of
``free'' quarks in this regime.

\begin{figure}
\doublefig
{\includegraphics[width=\textwidth]{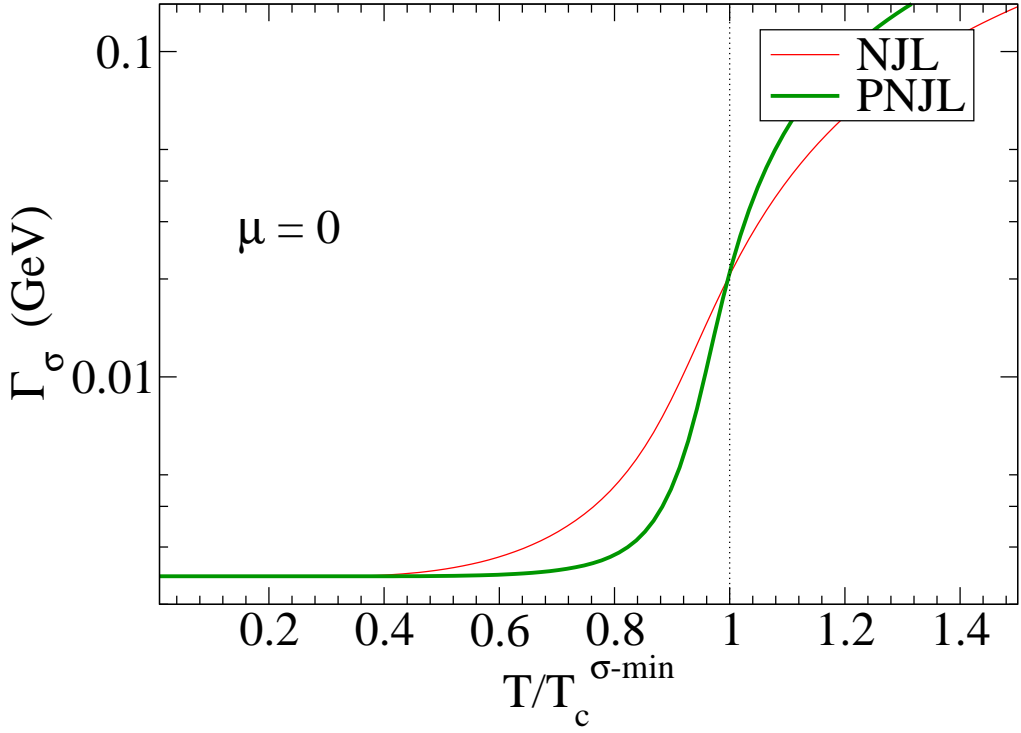}}
{\includegraphics[width=\textwidth]{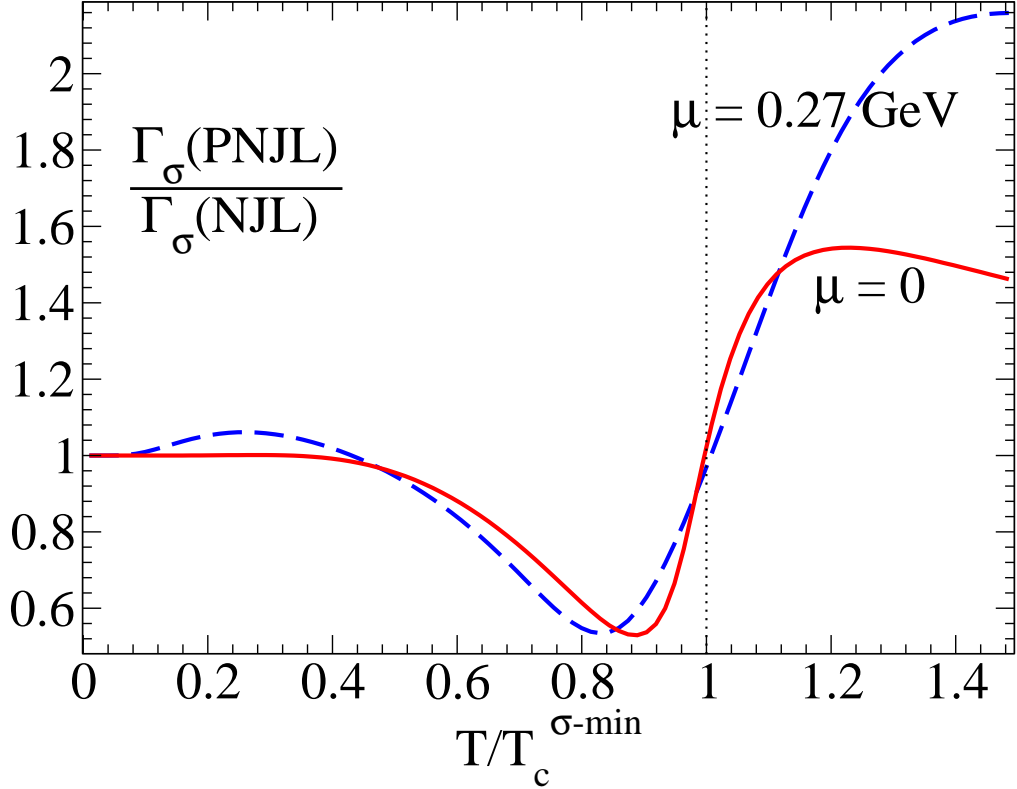}}
\caption{Comparison of the width of the $\sigma$ in NJL and PNJL as a
function of the reduced temperature $T/T_c$.
Top: absolute comparison at zero chemical
potential. Bottom: relative width
$\Gamma_\sigma(PNJL)/\Gamma_\sigma(NJL)$ at $\mu = 0$ and $\mu = 0.27$ GeV.
In the latter, a maximum
effect below $T_c$ ($T \simeq 0.87\, T_c$) of $\simeq 40$\% can be seen.}
\label{fig:WH}\label{fig:WRel}
\end{figure}

To quantitatively illustrate these features at zero chemical potential, in 
 {the upper panel of Fig.~\ref{fig:WH} we show the absolute
values of the $\sigma$ meson width in the NJL and in the PNJL models,
while in the lower panel we show the ratio between the widths
evaluated in the two models.} The width of the $\sigma$ evaluated in PNJL
is smaller than the NJL one below $\Tms$ and it is larger above
$\Tms$. This is an indication that in PNJL the decay channel
$\sigma\rightarrow \bar{q}q$ is reduced at low temperatures with
respect to the NJL case.  On the contrary, above $\Tms$ one can
interpret the larger width of the quarks bound into the meson as a
more efficient deconfinement effect.
In spite of the smallness of both absolute widths, the PNJL model
entails a reduction up to $\simeq 40$ \% for $\Gamma_\sigma(PNJL)$, a
step toward confinement: this can be seen in the lower panel of
Fig.~\ref{fig:WH}, where the relative width
$\Gamma_\sigma(PNJL)/\Gamma_\sigma(NJL)$ is displayed as a function of
the reduced temperature.

We also notice that, in both channels and at $\mu = 0$,
Eq.(\ref{eq:mesonmass}) no longer has a real solution for
$T=0.396$~GeV$=2.04 \Tcc$ in NJL and $T=0.422$~GeV$=1.65 \Tcc$ in
PNJL. These can be interpreted as the dissociation temperatures of the
model~; the faster (in relative units) occurrence of dissociation in
PNJL is in agreement with the larger width of the $\sigma$ meson at
high temperatures.

In Fig.\ref{fig:WRel} (bottom) we also report the ratio of widths at a finite
chemical potential, $\mu = 0.27$ GeV. Here the overall situation is qualitatively
similar to what happens at $\mu = 0$. However the curve shows some peculiar
features, overshooting one at small $T/T_c$ (where the absolute widths
are both, in any case, very small). This behavior does not lend
itself to an immediate physical interpretation since the critical point,
in the two models, differs not only for the value of the temperature
but also of the chemical potential. Hence the ``absolute'' value $\mu = 0.27$
GeV corresponds to different physical situations: a rescaling for the chemical
potential should be performed, but it goes beyond the scope of
the present work, since it involves a precise discussion of the phase
diagram of the PNJL model compared to the NJL one.

\section{Conclusions \label{discussion}} 
In the present work, we have investigated the properties of scalar and
pseudo-scalar mesons at finite temperature and quark chemical potential
in the framework of the Polyakov loop extended Nambu--Jona-Lasinio model.
This model has proven to be particularly successful in reproducing
two flavour QCD thermodynamics as obtained in lattice calculations
\cite{Ratti:2005jh}: the coupling of quarks to the Polyakov loop 
produces a statistical suppression of the one- and two-quark contributions
to the thermodynamics, thus remarkably improving the NJL model results at low 
temperatures. 

The present work was meant as a test of the PNJL model in the
mesonic sector. On the one hand, it was important to check whether the
role of pions as Goldstone bosons as well as the pion-$\sigma$ degeneracy in
the chirally restored phase are still satisfied after coupling quarks
to the Polyakov loop. On the other hand, it was interesting to
investigate whether the coupling to the Polyakov loop can cure some
problems of the ``classic'' NJL model description of mesons, such as
the unphysical width of the $\sigma$ meson for the process
$\sigma\rightarrow\bar{q}q$ in the chirally broken phase.

Finally we also intended to generalize the NJL formalism in order to 
embody the Polyakov loop coupled to quarks. This turned out to be 
particularly useful in the mesonic sector.
Indeed
we have shown the important results that PNJL calculations can be
directly deduced from NJL ones (not only for one loop calculations,
but to all orders) simply by a redefinition of the usual Fermi --
Dirac distribution function.

Our work shows a perfect agreement between the NJL and PNJL results
concerning the mesonic masses: in the high temperature phase, pions
and $\sigma$ tend to merge, thus displaying the correct pattern for
chiral symmetry restoration. In particular, the pions still survive as
bound states up to $T_{Mott}\simeq1.06\Tcc$, and their Goldstone boson
nature is still preserved in the chirally broken phase.

As far as the $\sigma$ meson is concerned, no true confinement is
observed in the model, since the unphysical width due to the decay
into a $\bar{q}q$ pair is still present in the PNJL model. This does
not come as a surprise, since no dynamical, self-coupled gluons are
embodied in the model Lagrangian.  In any case our results in
PNJL on the decay width improve slightly the NJL ones.

\begin{acknowledgments}
We thank J. Aichelin and W. Weise for stimulating discussions. 
One of the authors (A.B.) thanks the Fondazione Della Riccia for financial support
and ECT* for the warm hospitality during the first part of this work.
\end{acknowledgments}

\end{document}